\documentclass[structabstract]{aa}  
\usepackage{graphicx}
\usepackage{longtable}
\usepackage{lscape}
\usepackage{txfonts}
\usepackage{natbib}
\bibpunct{(}{)}{;}{a}{}{,}
\begin{document}
   \title{Rotating and infalling motion around the high-mass young stellar object Cepheus~A-HW2
            observed with the methanol maser at 6.7~GHz}

   \subtitle{}

   \author{K. Sugiyama
          \inst{\ref{inst1}},
          K. Fujisawa\inst{\ref{inst2},\ref{inst1}},
          A. Doi\inst{\ref{inst3},\ref{inst4}},
%T. Hirota\inst{\ref{inst5},\ref{inst6}},
          M. Honma\inst{\ref{inst5},\ref{inst6}},
          H. Kobayashi\inst{\ref{inst5},\ref{inst7}},
          Y. Murata\inst{\ref{inst3},\ref{inst4}},
          K. Motogi\inst{\ref{inst2}},\\
          K. Niinuma\inst{\ref{inst1}},
          H. Ogawa\inst{\ref{inst8}},
          K. Wajima\inst{\ref{inst9}},
          S. Sawada-Satoh\inst{\ref{inst7}},
          \and
          S.~P. Ellingsen\inst{\ref{inst10}}          
%\fnmsep\thanks{Just to show the usage
%          of the elements in the author field}
          }

   \institute{Graduate school of Science and Engineering,
              Yamaguchi University, 1677-1 Yoshida, Yamaguchi,
              Yamaguchi 753-8512, Japan\\
              \email{[koichiro;niinuma]@yamaguchi-u.ac.jp}
              \label{inst1}
         \and
             Research Institute for Time Studies, Yamaguchi University,
             1677-1 Yoshida, Yamaguchi, Yamaguchi 753-8511, Japan\\
             \email{[kenta;motogi]@yamaguchi-u.ac.jp}\label
             {inst2}
         \and
             The Institute of Space and Astronautical Science,
             Japan Aerospace Exploration Agency, 
             3-1-1 Yoshinodai,\\Chuou-ku, Sagamihara, Kanagawa 252-5210, Japan\\
             \email{[akihiro.doi;murata]@vsop.isas.jaxa.jp}
             \label{inst3}
         \and
             Department of Space and Astronautical Science,
             The Graduate University for Advanced Studies,
             3-1-1 Yoshinodai,\\Chuou-ku, Sagamihara, Kanagawa 252-5210, Japan
             \label{inst4}
         \and
             VERA Project, National Astronomical Observatory of Japan
             (NAOJ), 2-21-1 Osawa, Mitaka, Tokyo 181-8588, Japan\\
             \email{[mareki.honma;hideyuki.kobayashi]@nao.ac.jp}
             \label{inst5}
         \and
             Department of Astronomical Science, Graduate University
             for Advanced Studies, 2-21-1 Osawa, Mitaka,
             Tokyo 181-8588, Japan
             \label{inst6}
         \and
             Mizusawa VLBI Observatory, NAOJ,
             2-12 Hoshigaoka-cho, Mizusawa-ku, Oshu, Iwate 023-0861, Japan\\
             \email{satoko.ss@nao.ac.jp}
             \label{inst7}
         \and
             Department of Physical Science, School of Science,
             Osaka Prefecture University, 1-1 Gakuen-cho, Nakaku,\\
             Sakai, Osaka 599-8531, Japan\\
             \email{ogawa@p.s.osakafu-u.ac.jp}
             \label{inst8}
         \and
             Shanghai Astronomical Observatory,
             Chinese Academy of Sciences,
             80 Nandan Road, Shanghai, 200030, China\\
             \email{kwajima@shao.ac.cn}
             \label{inst9}             
         \and
             School of Mathematics and Physics, University of Tasmania,
             Private Bag 37, Hobart, Tasmania 7001, Australia\\
             \email{Simon.Ellingsen@utas.edu.au}
             \label{inst10}
%\thanks{The university of heaven temporarily does not
%                     accept e-mails}
             }

   \date{Received 12 February 2013; Accepted 03 December 2013}

% \abstract{}{}{}{}{} 
% 5 {} token are mandatory
 
  \abstract
  % context heading (optional)
  % {} leave it empty if necessary  
   {Proper motion observations of masers can provide information on dynamic motions on scales of a few milliarcsecond per year (mas~yr$^{-1}$) 
at radii of 100--1000~au scales from central young stellar objects (YSOs).
}
  % aims heading (mandatory)
   {The 6.7~GHz methanol masers are one of the best probes for investigations of the dynamics 
of high-mass YSOs, and in particular for tracing the rotating disk. We have measured the internal 
proper motions of the 6.7~GHz methanol masers associated with Cepheus A (Cep~A) HW2 
using Very Long Baseline Interferometery (VLBI) observations.
}
  % methods heading (mandatory)
   {We conducted three epochs of VLBI monitoring observations of the 6.7~GHz methanol masers 
in Cep~A-HW2 with the Japanese VLBI Network (JVN) over the period between 2006--2008. 
In 2006, we were able to use phase-referencing to measure the absolute coordinates 
of the maser emission with an accuracy of a few milliarcseconds. 
We compared the maser distribution with other molecular line observations that trace the rotating disk.
}
  % results heading (mandatory)
   {We measured the internal proper motions for 29 methanol maser spots,
of which 19 were identified at all three epochs and the remaining ten at only two epochs. 
The magnitude of proper motions ranged from 0.2 to 7.4~km~s$^{-1}$, with an average 
of 3.1~km~s$^{-1}$. 
Although there are large uncertainties in the observed internal proper motions of the 
methanol maser spots in Cep~A, they are well fitted by a disk that includes both rotation 
and infall velocity components.
The derived rotation and infall velocities at the disk radius of 680~au are 
$0.5 \pm 0.7$ and $1.8 \pm 0.7$~km~s$^{-1}$, respectively.
}
  % conclusions heading (optional), leave it empty if necessary 
   {Assuming that the modeled disk motion accurately represents the accretion disk around 
the Cep~A-HW2 high-mass YSO, we estimated the mass infall rate to be 
$3 \times 10^{-4} n_{8}~M_{\sun}~\mathrm{yr}^{-1}$ ($n_{8}$ is the gas volume density 
in units of $10^{8}~\mathrm{cm}^{-3}$).
The combination of the estimated mass infall rate and the magnitude of the fitted infall velocity 
suggests that Cep~A-HW2 is at an evolutionary phase of active gas accretion from the disk 
onto the central high-mass YSO.
The infall momentum rate is estimated to be 
$5 \times 10^{-4} n_{8}~M_{\sun}~\mathrm{yr}^{-1}~\mathrm{km~s}^{-1}$, which is larger 
than the estimated stellar radiation pressure of the HW2 object, supporting the hypothesis 
that this object is in an active gas accretion phase.
}


   \titlerunning{Infall motion of the Cep~A methanol masers at 6.7~GHz}
   \authorrunning{K. Sugiyama et al.}
   
   \maketitle
%
%________________________________________________________________

%%% Section 1. Introduction %%%
\section{Introduction}\label{section1}

In the last decade, high-resolution interferometric observations at submillimeter wavelengths 
have demonstrated the existence of rotating disks around high-mass young stellar objects 
(YSOs) \citep[e.g.,][]{Cesaro97,Cesaro99,Patel05}. 
These observations have shown that the spatial distribution of dust, or that of hot molecular 
core tracers (e.g., CH$_3$CN lines) are elongated perpendicular to the direction of a radio 
jet or large-scale outflow,
and that radial velocity gradients are observed along the direction of elongation. 
The recently developed capability to make near-infrared interferometric observations 
has been utilized to demonstrate the existence of a compact dusty disk ($\sim$20~au) 
around the high-mass YSO IRAS~13481$-$6124 \citep{Kraus10}. 
\citeauthor{Kraus10} found that the disk in IRAS~13481$-$6124 has a similar radial 
temperature gradient and scale for the dust-free region to those observed 
in low-mass star formation regions. 
A direct measurement of the distribution and 3-D motion of gas in a disk associated 
with a high-mass star represents the next step to be undertaken in studies of this type of object.

Molecular maser emission is potentially a useful tool for investigating the distribution and motion 
of gas around YSOs. 
The maser emission is typically compact (linear scales of 1--10~au), has a high brightness 
temperature ($>10^6$~K), and narrow line width ($<1$~km~s$^{-1}$). 
In particular, studies of the internal proper motion of individual maser spots 
(the compact emission centers detected at a given velocity channel) with Very Long Baseline 
Interferometery (VLBI) can provide directly information on the motion of the masing gas 
on scales of a few milliarcsecond per year (mas~yr$^{-1}$). 
Among the various common interstellar masers, the methanol masers at 6.7~GHz are 
perhaps the best probe of the gas motion around high-mass YSOs. 
This is because the 6.7~GHz methanol masers are only associated with high-mass star formation 
\citep{Minier03,Xu08}, the individual maser features have a long lifetime \citep{van05,Elli07}, 
and they are typically associated with star formation regions at an early evolutionary phase prior 
to the formation of an ultra-compact (UC) \ion{H}{II} region \citep{Walsh98,Minier05,Elli06}.
Interferometric imaging surveys of 6.7~GHz methanol masers 
\citep{Nor93,Phill98,Walsh98,Minier00,Sugi08a,Bart09,DeBui03,Dod04}
show that they are often associated with disks and shocks around high-mass YSOs.
Recently, \citet{Bart09} showed that nine of the 31 sources they observed ($\sim$30{\%}) 
have a ring-like morphology. 
They applied a disk model with rotation and expansion/contraction to the observed maser 
distribution and radial velocities and noted that, in general, the expansion/contraction 
velocity was larger than the rotation component. 
However, because the velocity was measured only along the line-of-sight (l.o.s.) the 
Bartkiewicz et al. models are poorly constrained due to geometrical ambiguity. 
Measuring the internal proper motions of the maser spots is required to derive the gas 
motion around the YSOs without ambiguity.

Rotational motion associated with a disk has been demonstrated by measuring 
the internal proper motions in a few methanol masers \citep{Sanna10a,Sanna10b,Mosca11}, 
while outflow motion has been measured in other methanol sources 
\citep{Rygl10,Sugi11,Matsu11,Sawada13}.
\citet{Goddi11} directly detected infall motion with a velocity of 5~km~s$^{-1}$ in the 
molecular envelope of the high-mass YSO AFGL~5142 MM-1 at a radius of 300~au.

The 6.7~GHz methanol masers associated with Cep~A-HW2 represent one of the best targets 
for measuring the rotation and infall motions through an accretion disk toward a high-mass YSO. 
The high-mass star-forming region Cep~A is located at a close distance of 0.70~kpc 
\citep{Mosca09,Dzib11}.
The HW2 object is the brightest source in the Cep~A complex contributing half of 
the total luminosity of the region $\sim 2.5 \times 10^{4} \,L_{\sun}$ 
\citep{Evan81,Rodri94,Hugh95,Garay96}.
The binding mass enclosed in HW2 is estimated to be $\sim 20 \,M_{\sun}$ 
\citep{Patel05,Jime09} and a fast, bipolar, highly collimated radio jet 
is being ejected from the HW2 YSO at an ejection velocity $\sim500$~km~s$^{-1}$, 
size $\sim 1500$~au, and position angle (PA) $\sim 45\degr$ \citep{Curi06}.
Observations of both dust and CH$_3$CN line emission made with the Submillimeter Array (SMA) 
show a rotating disklike structure elongated perpendicularly to the radio jet, 
and with a l.o.s. velocity gradient along the direction of elongation
\citep{Patel05}. This discovery has been confirmed through investigations of the spatial distribution 
and velocity gradients of NH$_3$, SO$_2$, and HC$_3$N emission \citep{Torre07,Jime07,Jime09}.

Observations of the 6.7~GHz methanol maser emission show an elliptical morphology, 
with the HW2 YSO at the center of the ellipse \citep{Sugi08b,Vlem10,Torst11a}.
The elliptical distribution of the masers is similar to that observed in the molecular 
and dust emission and the size of the ellipse ($\sim$1300--1400~au) is similar to 
that of the CH$_3$CN and the NH$_3$ disks. 
\citet{Torst11a} fitted a ring model with rotation and infall motions \citep{Usca08} 
to the observed methanol maser positions and l.o.s. velocities, 
and derived an infall velocity of $+1.3$~km~s$^{-1}$ and a rotation velocity of $+0.2$~km~s$^{-1}$.
\citeauthor{Torst11a} concluded that the infall component of velocity prevailed 
over the rotation component.
\citet{Vlem10} determined the 3-D magnetic field structure around HW2 and found 
the magnetic field to lie perpendicular to the molecular and dust disks, 
indicating that the magnetic field likely regulates the accretion onto the disk.

We have conducted VLBI monitoring observations of the 6.7~GHz methanol masers in Cep~A-HW2, 
using the Japanese VLBI Network \citep[JVN:][]{Fuji08} at three epochs spanning 779 days, 
to measure the maser internal proper motions.
We describe these observations and the details of data reduction in section~\ref{section2}. 
In section~\ref{section3}, we show the maser spatial distribution
and internal proper motions. Finally, we discuss the interpretation of the measured motion 
of the methanol masers in section~\ref{section4}.

%%% Section 2. Observations and data reduction %%%
\section{Observations and data reduction}\label{section2}

%%% Table 1. %%%
\begin{table*}[t]
\centering
\caption{Parameters of VLBI observations using the JVN for Cep~A.}
\label{tab:tab1}
\begin{tabular}{cclccccc}
\hline\hline
Epoch & Date and Time    & \multicolumn{1}{c}{Telescopes\tablefootmark{*}}
& $t_{\mathrm{on}}$ & 1~$\sigma$
& \multicolumn{2}{c}{Synthesized beam} &
$N_{\mathrm{spot}}$ \\ \cline{6-7}
      &          &
&                   &
& $\theta_{\mathrm{maj}} \times \theta_{\mathrm{min}}$ & PA &  \\
      &  (yyyy/mm/dd, UT) &
&   (hr)            & (Jy~beam$^{-1}$)
& (mas$\times$mas)  &  ($\circ$) & \\ \hline
1   & 2006/09/09, 14:30$-$22:00 & YM, UD, MZ, IG
& 2.8 & 0.06 & 9.4$\times$4.3 & $-$70 & 84 \\
2   & 2007/07/28, 14:00$-$22:00 & YM, UD, MZ, IR, IG
& 2.1 & 0.09 & 9.2$\times$4.3 & $-$43 & 92 \\
3   & 2008/10/25, 10:00$-$18:00 & YM, MZ, IR, OG, IG
& 1.6 & 0.12 & 7.0$\times$3.9 & $-$79 & 60 \\ \hline
\end{tabular}
\tablefoot{\footnotesize
Column~1: epoch number; Column~2: observational year/month/day,
and universal time;
Column~3: telescopes used; Column~4: total on-source time;
Column~5: image rms noise in a line-free channel;
Columns~6--7: FWHM of major and minor axes,
and position angle of synthesized beam made 
with natural weighting;
Column~8: number of detected maser spots.\\
\tablefoottext{*}{Telescope code  --- YM: Yamaguchi,
UD: Usuda, MZ: VERA-Mizusawa, IR: VERA-Iriki,
OG: VERA-Ogasawara, IG: VERA-Ishigaki.}
}
\end{table*}

We made the JVN observations at three epochs: 2006 September 9, 2007 July 28, 
and 2008 October 25. 
The results of the imaging from the first epoch were reported by \citet{Sugi08b}. 
The array for each epoch consisted of four or five 
of the following radio telescopes: Yamaguchi~32~m, Usuda~64~m, and VERA 
(four 20~m stations: Mizusawa, Iriki, Ogasawara, and Ishigaki). 
The telescopes used in each observation are listed in Table~\ref{tab:tab1} along with 
other relevant observational parameters, such as the duration of the observation and 
the size of the synthesized beam. 
The projected baselines ranged from 6~M$\lambda$ (Yamaguchi--Iriki) to 50~M$\lambda$ 
(Mizusawa--Ishigaki), corresponding to fringe spacings of 
34.4~mas and 4.1~mas at 6.7~GHz, respectively.

We made the observations using phase-referencing to determine the absolute position 
of the 6.7~GHz masers in Cep~A with an accuracy of a few milliarcseconds. 
We used the extragalactic continuum source J2302$+$6405 (2.19$ \degr$ from Cep~A) 
as the phase calibrator. 
We determined the coordinates of J2302$+$6405 to an accuracy of 0.62~mas 
in the third VLBA Calibrator Survey catalog \citep[VCS3:][]{Petrov05}. 
The observation strategy employed in all epochs was to alternate between the position of 
the Cep~A methanol masers and the phase calibrator with a cycle time of 5~min 
(2~min on Cep~A, 1.6~min on the continuum source and 1.4~min slewing between sources). 
The total on-source times were 2.8, 2.1, and 1.6 h for Cep~A and 0.7, 1.0, and 0.9 h 
for J2302$+$6405 in epochs 1 to 3, respectively. 
The absolute coordinates obtained at the first epoch were previously reported by \citet{Sugi08b}. 
The nearby radio continuum sources J2322$+$5057 and J0102$+$5824, 
and the strong continuum sources 3C454.3 and 3C84, selected from the International Celestial 
Reference Frame \citep[ICRF:][]{Ma98,Fey04}, were also observed every 1.5 hour for clock 
(delay and rate) and bandpass calibration, respectively.

For the first epoch, we observed left-circular polarization (LCP) at the Yamaguchi and Usuda stations, 
while received linear polarization at the Mizusawa and Ishigaki stations. 
At the following epochs, we observed LCP at Yamaguchi, Usuda, Mizusawa, and Ishigaki stations, 
while received linear polarization at the Ogasawara and Iriki stations. 
Employing a bandwidth of 32~MHz, we recorded the data to magnetic tape 
using the VSOP-terminal system at a data rate of 128~Mbps with 2-bit quantization, 
and correlated at the Mitaka FX correlator \citep{Shiba98}. 
From the recorded 32~MHz bandwidth, we correlated 2~MHz (6668--6670~MHz) and 
4~MHz (6666--6670~MHz) sub-bands with 512 and 1024 spectral channels for epochs 1 
and epochs 2/3, respectively. Both spectral setups yielded a channel spacing of 0.18~km~s$^{-1}$.
The VLBI data were reduced using the Astronomical Image Processing System 
\citep[AIPS:][]{Grei03} using the same procedure described in \citet{Sugi08a}, 
which includes special amplitude calibration for different polarization correlations 
(circular/linear, linear/linear).
The image rms noise (1~$\sigma$) in a line-free channel was 0.06, 0.09, and 0.12~Jy~beam$^{-1}$ 
for epochs 1, 2, and 3, respectively, as listed in Table~\ref{tab:tab1}.
Phase-referencing was successful only at the first epoch 
with the best image sensitivity.

\begin{figure}[htbp]
\begin{center}
\includegraphics[width=77mm,clip]{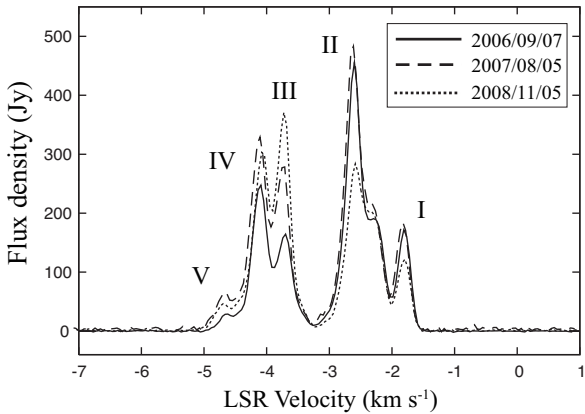}
\end{center}
\caption{{\footnotesize
Spectra of the 6.7~GHz methanol maser in Cep~A obtained with the Yamaguchi 32-m radio telescope. 
Solid, dashed, and dotted lines show the spectra observed near in time to the VLBI 
epochs 1, 2 and 3, respectively. The Roman numerals I-V identify different spectral features.
}}
\label{fig1}
\end{figure}

We performed single-dish observations using the Yamaguchi 32~m telescope 
to enable absolute flux calibration of the VLBI observations.
We made the single-dish observations on 2006 September 7, 2007 August 5, 
and 2008 November 5, respectively, 
each of them separated by at most ten days from the corresponding VLBI epoch.
The flux density calibration template spectrum for each epoch is shown in Figure~\ref{fig1}. 
The channel spacing of these single-dish observations was four times higher than 
that of the VLBI data, i.e., 0.044~km~s$^{-1}$. 
In 2007, the 6.7~GHz methanol masers in Cep~A were observed to show rapid variations 
in flux density within a 30 day period \citep{Sugi08b}. 
Assuming that flux density variations of up to $\sim $50{\%} within 30~days are possible, 
we estimate the absolute flux density calibration for the VLBI observations to be accurate to 20--30{\%}.
The estimated uncertainty is calculated from the accuracy of the absolute flux calibration 
for the single-dish, at 15--20{\%}, and the expected maximum flux variation 
within $\sim$10 days, at 15--20{\%}.
For each epoch, the amplitude calibration was performed 
by fitting the telescope total-power spectra to the single-dish template spectrum.

We used the brightest 6.7~GHz maser emission at a local standard of rest (LSR) velocity 
of $-$2.60~km~s$^{-1}$ as the reference maser spot \citep{Sugi08b} in fringe fitting, 
and applied the phase solution from this spectral channel to the visibilities of all velocity channels. 
We searched for maser emission over an area of $4.0\arcsec \times 4.0\arcsec$ using 
the Difmap software \citep{Sheph97} by model fitting the visibility data with  point sources 
and iteratively self-calibrating. 
We determined the peak positions and intensities of maser spots by fitting an elliptical 
Gaussian brightness distribution to the emission pattern in each spectral channel 
using the task JMFIT in AIPS.
We used one or more independent Gaussian components
with about the same dimensions as the synthesized beam 
to prevent the fitted Gaussians from being significantly larger than the beam.
Maser components are considered real if detected with signal-to-noise ratio (SNR) $\geq 5$ at similar (within the beam FWHM) positions in two or more consecutive channels.
Maser structure is not resolved
even on the longer baselines, and the correlated flux density recovered from the final images contains more than 80{\%} of the single-dish flux density.

%%% Section 3. Results %%%
\section{Results}\label{section3}
\subsection{Spatial distributions}\label{section3-1}

We detected 84, 92, and 60 methanol maser spots in epochs 1, 2, and 3, respectively. 
The total number of observed, distinct maser spots is 131 if we take the fact that 
many spots are observed at more than one epoch into account. 
The spot peak intensities range from $\sim$0.4 to 92.2~Jy~beam$^{-1}$.
The lower number of detected spots in our first epoch run, compared with the observation 
by \citet{Sugi08b}, is due to our higher detection threshold to select spots strong enough 
to allow for an accurate measure of the proper motion.
Maser variability, as evidenced by the single-dish spectra, can explain the different 
number of maser detections over the three observing epochs.
The parameters of each maser spot, as derived by the Gaussian fit, are summarized in 
Table~\ref{longtab1}, which lists: the ID number of each maser spot and cluster; 
the relative positional offset with respect to the reference spot (ID 53) (and estimated uncertainty); 
the spot LSR velocity; the internal proper motion (and estimated uncertainty) and the tangential 
velocity (converted on the basis of a source distance of 0.70~kpc); 
and the peak intensity at each epoch.

\begin{figure}[htbp]
\begin{center}
\includegraphics[width=77mm,clip]{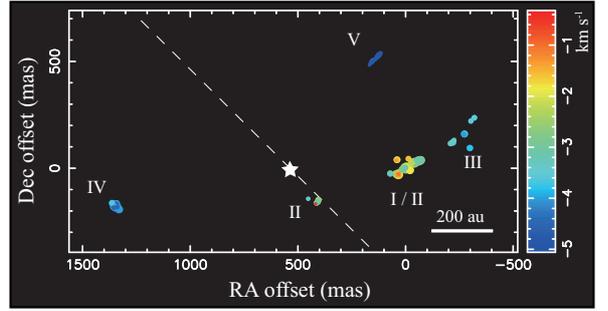}
\end{center}
\caption{{\footnotesize
Spatial distribution of the 6.7~GHz methanol maser spots in Cep~A (\textit{filled circles}). 
The size of the circles are proportional to the peak intensity on a logarithmic scale, 
and the color indicates the l.o.s. velocity (color-velocity conversion code shown 
on the right of the plot). 
The spatial scale is shown by the ruler at the bottom-right corner. 
The Roman numerals identify the maser clusters emitting 
the spectral features shown in Figure~\ref{fig1}. 
The star symbol indicates the peak of the 43~GHz continuum emission, 
which is known to a positional accuracy of 10~mas \citep{Curi06}. 
The dashed line shows the direction of the high-velocity collimated radio jet \citep{Curi06}. 
The origin of the map corresponds to the position of the reference 6.7~GHz methanol maser 
emission at LSR velocity of $- 2.60$~km~s$^{-1}$ 
[$\alpha (\mathrm{J2000.0}) = 22^{\mathrm{h}} 56^{\mathrm{m}} 17^{\mathrm{s}}.9042$, 
$\delta (\mathrm{J2000.0}) = +62 \degr 01 \arcmin49{\arcsec}.577$], 
as determined by \citet{Sugi08b}.
}}
\label{fig2}
\end{figure}

The spatial distribution of the 6.7~GHz methanol maser spots in Cep~A is shown in 
Figure~\ref{fig2} and includes all 131 spots detected in at least one epoch 
and listed in Table~\ref{longtab1}. 
The absolute coordinates of the reference maser emission at $-$2.60~km~s$^{-1}$ was 
measured by \citet{Sugi08b} to be 
$\alpha (\mathrm{J2000.0}) = 22^{\mathrm{h}} 56^{\mathrm{m}} 17^{\mathrm{s}}.9042$, 
$\delta (\mathrm{J2000.0}) = +62 \degr 01 \arcmin 49{\arcsec}.577$, 
with a positional uncertainy of less than 1~mas. 
The reference maser spot (at the origin of Fig.~2) was the brightest at all epochs. 
Our result is consistent with previous published images of the 6.7~GHz methanol masers 
in Cep~A observed by \citet{Sugi08b}, \citet{Vlem10}, and \citet{Torst11a}. 
The methanol maser spots in Cep~A are arranged in a number of isolated clusters, 
which we have labeled I to V using the same designation as \citet{Sugi08b}.

The 6.7~GHz maser spots in Cep~A are distributed along a curved line with size 
$\sim$1900~mas, corresponding to $\sim$1400~au at a distance of 0.70~kpc. 
The peak of the 43~GHz continuum emission, corresponding to the likely location 
of the YSO \citep{Curi06}, is located near the center of the curved line. 
The elongation of the maser line is nearly perpendicular to the sky-projected axis 
of the high-velocity collimated radio jet.
The position, size, and orientation of the 6.7~GHz maser distribution agrees with that 
of the molecular disk observed in CH$_3$CN and NH$_3$ lines \citep{Patel05,Torre07}, 
and the velocity range covered by the 6.7~GHz masers is similar to that of these lines. 
Although we found a simple velocity gradient along the major axis of elongation 
is not observed in the maser emission. 
We fitted the observed distribution of maser spots at each epoch with an ellipse. 
The parameters of the fit for each epoch are summarized in Table~\ref{tab:tab2}. 
The fitted parameters are the center of the ellipse relative to the reference position 
$(\Delta \alpha, \Delta \delta)$, the length of the semi-major axis of the ellipse $a$, 
the position angle of the major axis PA$_{\mathrm{mj}}$ 
(north is $0 \degr$, and counterclockwise is positive), and the inclination angle $i$.
The inclination angle is determined by assuming that the maser distribution is circular 
and the observer is looking at it obliquely (face-on corresponds to $i = 0 \degr$). 
The fit ellipse parameters are consistent with those derived 
from observations with the Multi-Element Radio Linked Interferometer Network 
of \citet{Vlem10} and with the European VLBI Network of \citet{Torst11a}, 
who found $a \sim 650$~au, PA$_{\mathrm{mj}} = 102 \degr$, $i = 71 \degr$ and 
$a \sim 678$~au, PA$_{\mathrm{mj}} = 99 \degr$, $i = 67.5 \degr$, respectively.

%%% Figure 1. %%%

%%% Figure 2. %%%

%%% Table 2. %%%
\begin{table}[htbp]
\caption{Parameters of the ellipse fitted to the observed maser distribution at each epoch.}\label{tab:tab2}
\centering
\begin{tabular}{crrrrr}
\hline\hline
Epoch & \multicolumn{2}{c}{Coordinates} & \multicolumn{1}{c}{$a$} & \multicolumn{1}{c}{PA$_{\mathrm{mj}}$} & \multicolumn{1}{c}{$i$} \\ \cline{2-3}
         & \multicolumn{1}{c}{$\Delta \alpha$} & \multicolumn{1}{c}{$\Delta \delta$} &    &  &  \\
         & \multicolumn{1}{c}{(mas)}       &  \multicolumn{1}{c}{(mas)} & \multicolumn{1}{c}{(au)} & \multicolumn{1}{c}{($ \degr$)}
& \multicolumn{1}{c}{($ \degr$)} \\ \hline
1   & $+$436 & $+$118 &  680 & 110 & 73 \\
2   & $+$385 & $+$125 &  756 & 114 & 75 \\
3   & $+$378 & $+$141 &  720 & 112 & 74 \\
\hline
\end{tabular}
\tablefoot{
Column~1: observational epoch;
Columns~2-3: east and north offset of the center of the ellipse relative to the maser reference position
[$\alpha (\mathrm{J2000.0}) = 22^{\mathrm{h}} 56^{\mathrm{m}} 17^{\mathrm{s}}.9042$, 
$\delta (\mathrm{J2000.0}) = +62 \degr 01 \arcmin49{\arcsec}.577$];
Columns~4-6: length and position angle (north is $0 \degr$, and counterclockwise is positive) 
of the semi-major axis, and inclination of the ellipse (face-on is $0 \degr$).
}
\end{table}

\subsection{Proper motion}\label{section3-2}
To establish the correspondence of maser spots observed at all the three epochs, 
we set the following criteria:
i)~LSR velocity to be the same to within the velocity width of a single spectral channel 
(0.18~km~s$^{-1}$),
ii)~relative positions of the three epochs to differ by less than 9.6~mas (corresponding to 
$\sim$15~km~s$^{-1}$ at 0.70~kpc, which is equivalent to three times the LSR velocity range),
iii)~stable intensity for the maser spectrum consisting of a maser spot and its nearby 
companions (within 10~mas) to avoid the ``Christmas tree effect" (see below),
iv)~spots assumed to move approximately along a straight line in both RA and Dec directions.
The purpose of the third criterion listed above is to ensure that the proper motions are real, 
and not due to the ``Christmas tree effect," where changes in the relative intensity of nearby 
maser spots with similar LSR velocity mimic proper motion. 
However, in the case where the derived proper motions 
are not parallel with the line connecting the nearby maser spots, 
we accept these proper motions as real gas motions even 
if the third criterion is not satisfied.
Proper motions of spots identified in only two of the three epochs
are considered only for spots 
persistent over consecutive epochs (i.e., between the first and second epoch 
and the second and third epoch, but not between the first and third epoch).
The maximum allowed tangential velocity of 15~km~s$^{-1}$ fixes the maximum position 
change between the epochs to 4.0~mas for maser spots observed only 
in the first and second epoch, and 5.6~mas for those observed only 
in the second and third epoch. 
After identifying all spots that satisfy the maximum tangential velocity two-epoch criteria, 
we compared their amplitude and direction of motion with that of nearby, 
three epoch persistent, linearly moving spots.
Only two-epoch spots, which have 
proper motions similar (both in amplitude and direction) to those of 
nearby three-epoch spots, were included in the final analysis. 
In the case of cluster V, the two-epoch proper motions were accepted if the amplitude of 
the tangential velocities were similar to those of three-epoch persistent, 
linearly moving spots in other clusters. 
Most of the accepted proper motions in cluster V move toward E-SE.
On the basis of these criteria, 19 maser spots were found to be persistent over the three 
epochs, six spots between the first and the second epoch, and four spots between the 
second and third epoch.
In total, we have 29 persistent maser spots and of these nine are associated with cluster I/II 
(located near the origin),
three with cluster III, seven with cluster IV, and ten with cluster V.

\begin{figure*}[htbp]
\begin{center}
\includegraphics[width=140mm,clip]{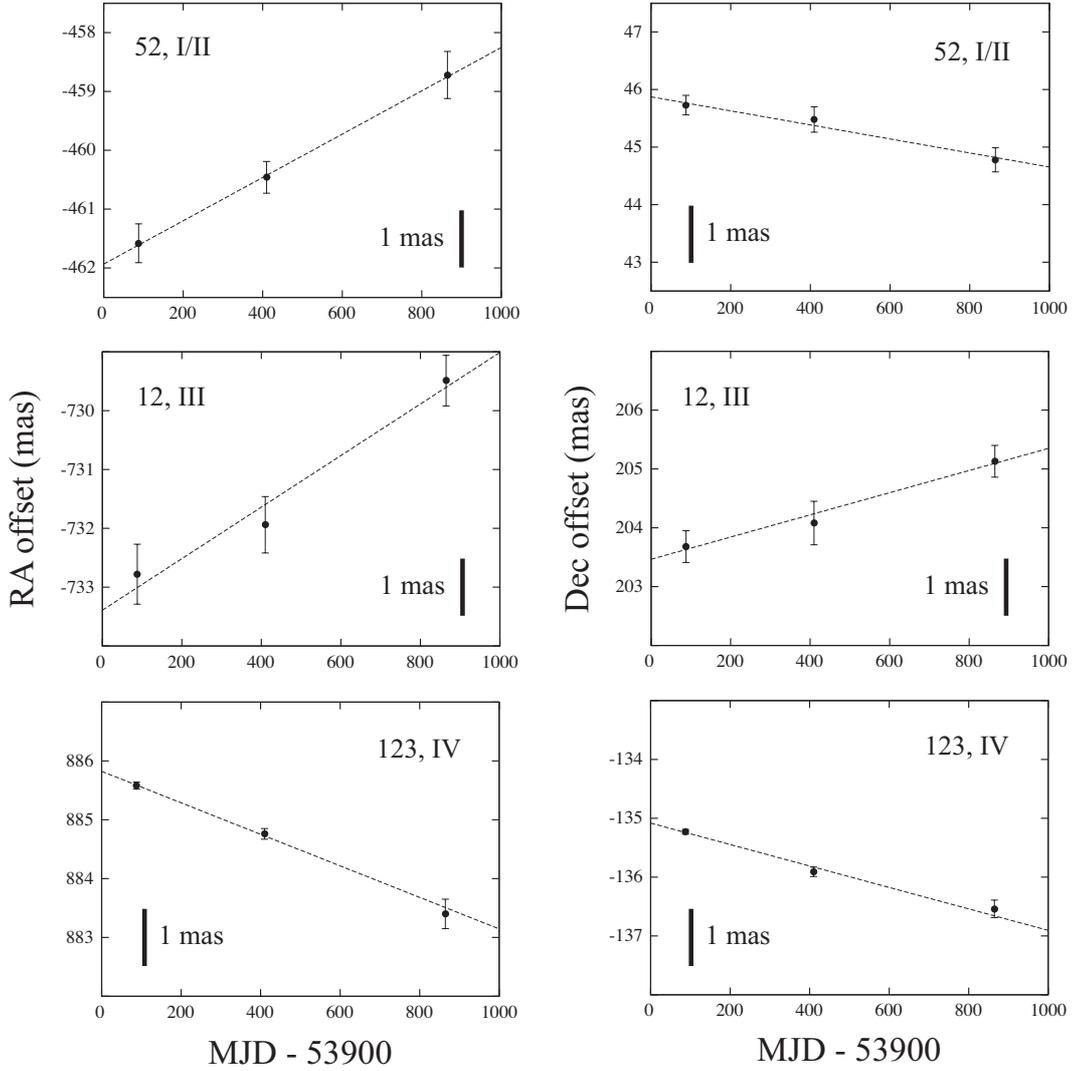}
\end{center}
\caption{{\footnotesize
Linear fits of the change with time (MJD)
of the relative positional offset
with respect to the barycenter
in right ascension (left-panel) and declination (right-panel).
Numbers and the Roman numerals 
identify the spot ID and the cluster as given in Table~\ref{longtab1}, respectively.
The dotted line indicates the linear fit.
}}
\label{fig3}
\end{figure*}

\begin{figure*}[htbp]
\begin{center}
\includegraphics[width=160mm,clip]{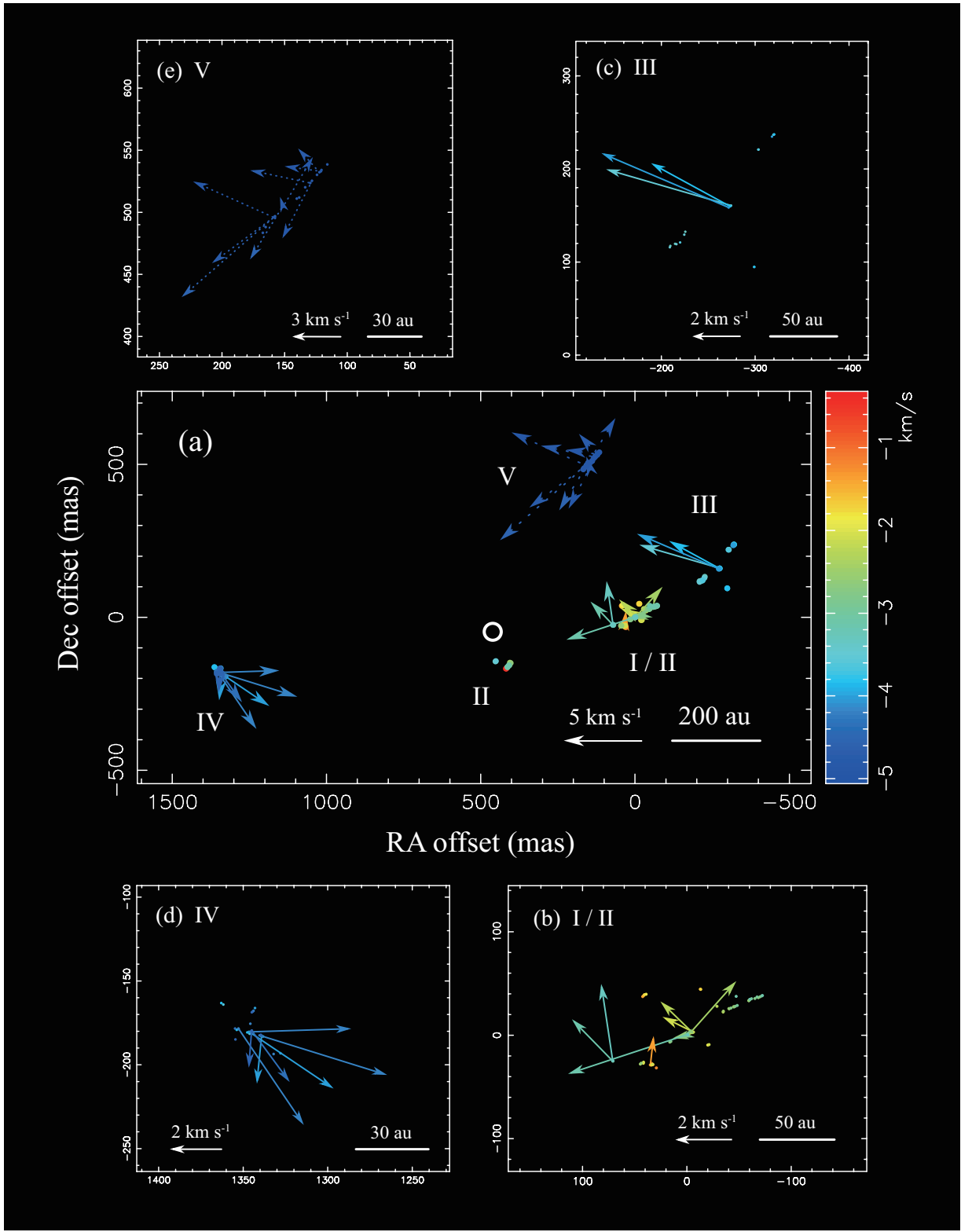}
\end{center}
\caption{{\footnotesize
Proper motions of the 6.7~GHz methanol maser emission in Cep~A relative to the barycenter
(see Section~\ref{section3-2}).
Arrows show the direction of the measured proper motions 
(solid line: using three epochs, dotted line: only two epochs).
Arrow length is proportional to the magnitude of the tangential velocity. 
The spatial and velocity scales are shown at
the lower-right corner in each panel.  The filled circles show the position of the maser spots.
The open circle indicates the barycenter of epoch 1.
(a)~Entire map,
(b)-(e)~Close-up maps for the cluster I/II, III, IV, and V, respectively.
}}
\label{fig4}
\end{figure*}

Internal proper motions are measured relative to the barycenter, defined as the average 
position of the 19 spots persistent over the three epochs.
The barycenter position relative to the reference maser spot at 
$V_{\mathrm{lsr}} = -2.60$~km~s$^{-1}$ 
is estimated to be [$+$461.01, $-$45.05]~mas, [$+$461.02, $-$45.39]~mas, 
and [$+$460.71, $-$44.97]~mas in epochs 1, 2, and 3, respectively.
The subtraction of the average position in each epoch is effective in removing random 
errors owing to the changes in the reference spot structure.
Proper motions are estimated by fitting a straight line to the positional offsets with respect to time.
Linear fits for representative maser spots in each cluster are shown in Figure~\ref{fig3}
to demonstrate the general quality of the fits.
The measured internal proper motions are given 
in columns 8-13 in Table~\ref{longtab1}.
The tangential velocities $V_{\mathrm{amp}}$ ($= \sqrt{V_{x}^{2} + V_{y}^{2}}$) vary 
in the range of $0.2 - 7.4$~km~s$^{-1}$ with an average of 3.1~km~s$^{-1}$. 
The measured internal proper motion vectors are shown in Figure~\ref{fig4}.
The barycenter of epoch 1 is indicated by an open circle in this figure.
This figure shows that the observed proper motions have a large (apparently random) scatter, 
particularly for cluster I/II, with no coherent motions readily apparent on these scales.  

\begin{figure*}[htbp]
\begin{center}
\includegraphics[width=125mm,clip]{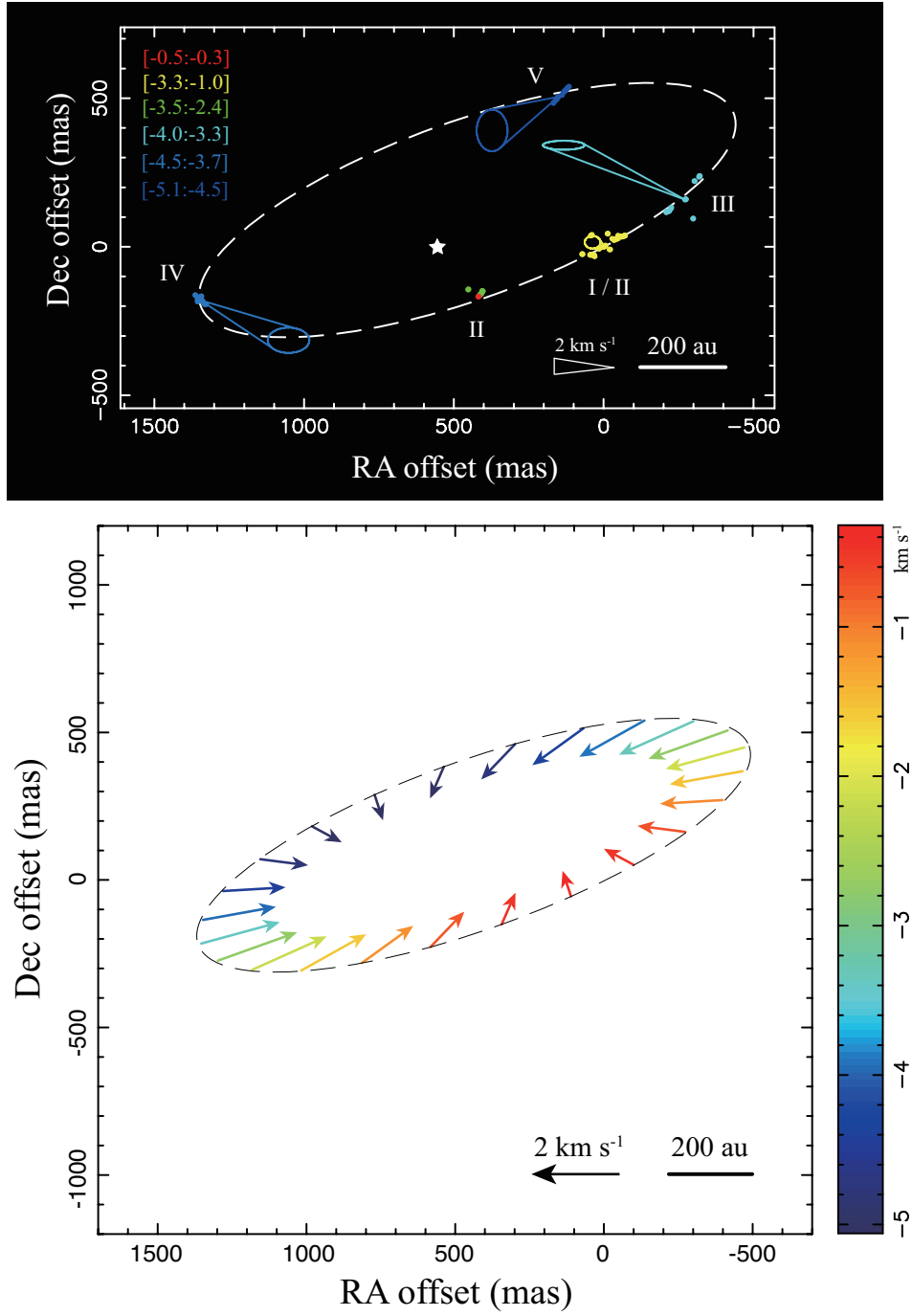}
\end{center}
\caption{{\footnotesize
\textit{Upper panel}: Internal proper motions averaged over all the persistent spots 
in each maser cluster, as listed in Table~\ref{tab:tab3}.
Cones show the averaged proper motions in each cluster 
(the aperture of the cone corresponds to the uncertainty).
The origin of each cone is located at the brightest spot in each cluster. 
The star represents the location of the 43~GHz continuum peak (as in Figure~\ref{fig2}). 
The dashed ellipse corresponds to the fitted elliptical structure for the methanol maser emission. 
Colors indicate the LSR velocity of maser clusters, with the range of LSR velocities 
within a cluster ([min:max]) shown in the top-left corner.
\textit{Lower panel}: Internal proper motions predicted from the fitted disk model 
including rotation and infall (see Section~\ref{section4-1}). 
The horizontal and vertical axes give the RA and Dec coordinates.
The spatial scale is indicated by the bar at the bottom-right corner of the plot. 
Colors indicate the LSR velocity of the maser emission located in the disk, 
using the same color velocity conversion code (shown on the right of the panel)
as in the upper panel.
The dashed ellipse is identical to that shown in upper panel.
}}
\label{fig5}
\end{figure*}

Taking the average of the proper motions of spots belonging to maser clusters I-V, 
we derived the maser cluster proper motions listed in Table~\ref{tab:tab3}, 
and shown in the upper-panel of Figure~\ref{fig5}. 
All the vectors, except that for cluster I/II, 
show rotation in a counterclockwise direction around the ellipse center.
In addition, all the cluster proper motion vectors are directed toward the inside of the maser ellipse.

%%% Table 3. %%%
\begin{table}[htbp]
\caption{Tangential velocities for the 6.7~GHz masers in Cep~A averaged over each maser cluster.}\label{tab:tab3}
\centering
\begin{tabular}{ccccc}
\hline\hline
Cluster & $V_{lsr}$           & $V_{x}$                  & $V_{y}$                  & $N_{\mathrm{spot}}$ \\
           & (km~s$^{-1}$)     & (km~s$^{-1}$)         &  (km~s$^{-1}$)        &                              \\ \hline
I/II       &  [$-$3.3, $-$1.0] & $+0.4 \pm 0.3$  & $+0.2 \pm 0.2$      &  9  \\
II          &  [$-$0.5, $-$0.3] & $\cdots$  & $\cdots$     &   $\cdots$ \\
           &   [$-$3.5, $-$2.4] & $\cdots$  & $\cdots$     &   $\cdots$ \\
III        &   [$-$4.0, $-$3.3] & $+4.0 \pm 0.7$  & $+1.8 \pm 0.2$     &   3  \\
IV        &    [$-$4.5, $-$3.7]  & $-2.9 \pm 0.7$  & $-1.3 \pm 0.4$     &    7  \\
V          &   [$-$5.1, $-$4.5]  &  $+2.2 \pm 0.5$  & $-1.1 \pm 0.7$     &    10  \\
\hline
\end{tabular}
\tablefoot{
Column~1: label of maser cluster ;
Column~2: range of l.o.s. velocity in the cluster ([min, max]) ;
Columns~3-4: averaged tangential velocity (toward east and north)
weighted by the error of each spot's proper motion
with uncertainty estimated as a weighted standard deviation;
Column~5: number of maser spots used to determine the average proper motion of cluster.\\
}
\end{table}

%%% Figure 3. %%%

%%% Figure 4. %%%

%%% Figure 5. %%%

%%% Section 4. Discussion %%%
\section{Discussion}\label{section4}

\subsection{Rotating and infalling motions around Cep~A-HW2}\label{section4-1}

%%% Table 4. %%%
\begin{table}[htbp]
\caption{Comparison of the ellipse fitted to the first epoch maser data and the CH$_3$CN molecular disk}
\label{tab:tab4}
\centering
\begin{tabular}{lccccc}
\hline\hline
         & \multicolumn{2}{c}{Coordinates (J2000.0)} & \multicolumn{1}{c}{$a$} & \multicolumn{1}{c}{PA$_{\mathrm{mj}}$} & \multicolumn{1}{c}{$i$} \\ \cline{2-3}
         & \multicolumn{1}{c}{RA} & \multicolumn{1}{c}{Dec} &    &  &  \\
         & \multicolumn{1}{c}{($^{\mathrm{h}}~^{\mathrm{m}}~^{\mathrm{s}}$)}    &  \multicolumn{1}{c}{(${\degr}~{\arcmin}~{\arcsec}$)}
& \multicolumn{1}{c}{(au)} & \multicolumn{1}{c}{($ \degr$)}
& \multicolumn{1}{c}{($ \degr$)} \\ \hline
Maser        & 22 56 17.97 & $+$62 01 49.7 &  680 & 110 & 73 \\
CH$_3$CN  & 22 56 17.96 & $+$62 01 49.6 &  560 & 124 & 68 \\
\hline
\end{tabular}
\tablefoot{
Column~1: Maser ellipse (fitted to 1st epoch data) or CH$_3$CN disk \citep{Patel05};
Columns~2-3: Absolute coordinates of the ellipse center;
Columns~4-6: radius of semi-major axis, position angle (north is $0 \degr$, and counterclockwise is positive),
and inclination of the ellipse or the disk (face-on is $0 \degr$).
}
\end{table}

The ellipse fitted to the methanol maser emission has very similar parameters to the disk traced 
by the CH$_3$CN emission \citep{Patel05}, although the radius of the maser ellipse is slightly 
larger than that of the CH$_3$CN disk (see Table~\ref{tab:tab4}).
This close agreement supports the hypothesis that the 6.7~GHz methanol masers around 
Cep~A-HW2 are associated with the CH$_3$CN and NH$_3$ disks.

We have fit a model of a rotating and expanding/contracting disk to the positions 
and 3-D velocities of the 6.7~GHz masers.
The model was applied to the 29 maser spots for which we measured the internal proper motions, 
under the assumption that all of the spots lie on a circular ring. 
The rotation velocity $V_{\mathrm{rot}}$, the expansion velocity $V_{\mathrm{exp}}$, 
and the systemic velocity $V_{\mathrm{sys}}$ of the model were related to the observed 
velocities as follows:
\begin{eqnarray*}
 V^{\mathrm{calc}}_{x'} &=& V_{\mathrm{rot}} \sin \theta + V_{\mathrm{exp}} \cos \theta \\
 V^{\mathrm{calc}}_{y'} &=& - (V_{\mathrm{rot}} \cos \theta - V_{\mathrm{exp}} \sin \theta) \cos i \\
 V^{\mathrm{calc}}_{z} &=& - (V_{\mathrm{rot}} \cos \theta - V_{\mathrm{exp}} \sin \theta) \sin i + V_{\mathrm{sys}}
\label{equ1}
\end{eqnarray*}
where $V^{\mathrm{calc}}_{x'}, V^{\mathrm{calc}}_{y'}$ are the tangential velocities
along the ellipse major and minor axis
and $V^{\mathrm{calc}}_{z}$ is the l.o.s. velocity, $\theta$ is the angle between the position 
vector to the spot and the ellipse major axis, and $i$ the inclination angle of the disk. 
The factors $\cos \theta$ and $\sin \theta$ can be expressed with $x' / a$ and $y' / (a \cdot \cos i)$, 
where $a$ is the semi-major axis 
and $x'$ and $y'$ are the coordinates along the major and minor axis of the ellipse, respectively.
The parameters $x'$ and $y'$ are given by:
\begin{eqnarray*}
x' &=& (y - y_{0}) \cos \mathrm{PA}_{\mathrm{mj}} + (x - x_{0}) \sin \mathrm{PA}_{\mathrm{mj}} \\
y' &=& (y - y_{0}) \sin \mathrm{PA}_{\mathrm{mj}} - (x - x_{0}) \cos \mathrm{PA}_{\mathrm{mj}}
\label{equ2}
\end{eqnarray*}
where $x$ and $y$ are the position offsets of the maser relative to the reference spot 
along the RA and Dec axis, respectively, $x_{0}$ and $y_{0}$ are the position offsets 
of the center of the ellipse and $\mathrm{PA}_{\mathrm{mj}}$ is the position angle 
of the ellipse major-axis.
Similar conversion formulae apply 
for the tangential velocity components as follows:
\begin{eqnarray*}
V_{x'} &=& V_{y}  \cos \mathrm{PA}_{\mathrm{mj}} + V_{x}  \sin \mathrm{PA}_{\mathrm{mj}} \\
V_{y'}  &=& V_{y}  \sin \mathrm{PA}_{\mathrm{mj}} - V_{x}  \cos \mathrm{PA}_{\mathrm{mj}}
\label{equ3}
\end{eqnarray*}
where $V_{x}$ and $V_{y}$ are the components of the tangential velocity 
along the RA and Dec axis, respectively.

The ellipse parameters $\Delta \alpha$ ($= x_{0}$), $\Delta \delta$ ($y_{0}$), $a$, 
PA$_{\mathrm{mj}}$, and $i$ are derived and fixed from the first epoch data (Table~\ref{tab:tab2}). 
The model parameters are derived 
by taking the partial derivative of Equation~(\ref{equ4}) with respect to 
$V_{\mathrm{rot}}$, $V_{\mathrm{exp}}$, $V_{\mathrm{sys}}$, 
respectively, and solving for when it is set equal to 0: 
\begin{eqnarray}
	 \chi^{2} = \sum^{N}_{j=1} w_{j} \bigl( (V_{x'j} - V^{\mathrm{calc}}_{x'j})^{2}
                    + (V_{y'j} - V^{\mathrm{calc}}_{y'j})^{2} + (V_{zj} - V^{\mathrm{calc}}_{zj})^{2}   \bigr)~.
\label{equ4}
\end{eqnarray}
In Equation~(\ref{equ4}), 
$N$ is the number of maser spots, $V_{x'j}$ and $V_{y'j}$ are 
the components of the tangential velocity of the maser spot with ID $j$ 
(projected along the ellipse major and minor axis), 
$V_{zj}$ is the LSR velocity, and $w_{j}$ is a weighting factor calculated 
as $1 / (\sigma^{2}_{xj} + \sigma^{2}_{yj} + \sigma^{2}_{zj})$. 
The factors $\sigma_{xj}, \sigma_{yj}$ are the estimated uncertainties in the tangential 
velocity components and $\sigma_{zj}$, the uncertainty in the LSR velocity, 
was set to the spectral-channel spacing (0.18~km~s$^{-1}$).

We derived the best-fit parameters from the data $V_{\mathrm{rot}} = + 0.5 \pm 0.7$, 
$V_{\mathrm{exp}} = - 1.8 \pm 0.7$ and $V_{\mathrm{sys}} = - 4.1 \pm 0.7$~km~s$^{-1}$, respectively. 
The negative expansion velocity indicates infall.
The uncertainties for each fit parameter were estimated by finding the variation from the best-fit value
for which the $\chi^{2}$ value increased by $\sim$10{\%}.
As a test of the reliability of the barycenter as the reference for the proper motions,
we redid the fit, including among the derived model parameters the tangential velocity
of the reference maser spot, and using spot velocities relative to the reference maser 
spot as inputs to the model.
From this test, we are able to estimate 
that the fitted velocity of the barycenter is zero within an uncertainty of $\pm 0.3$~km~s$^{-1}$. 
This uncertainty is smaller than the model-fit error, which confirms that the barycenter is a suitable
reference system for the proper motions.

We applied the same model to the cluster-averaged proper motions listed in Table~\ref{tab:tab3},
and obtained best-fit parameters $V_{\mathrm{rot}} = + 0.3 \pm 0.6$, 
$V_{\mathrm{exp}} = - 2.0 \pm 0.6$, $V_{\mathrm{sys}} = - 4.4 \pm 0.6$~km~s$^{-1}$, 
which are consistent with those derived using all 29 spots within the estimated uncertainties.
In the following discussion, we use the best-fit parameters derived with the 29 spots.

\begin{figure*}[htbp]
\begin{center}
\includegraphics[width=150mm,clip]{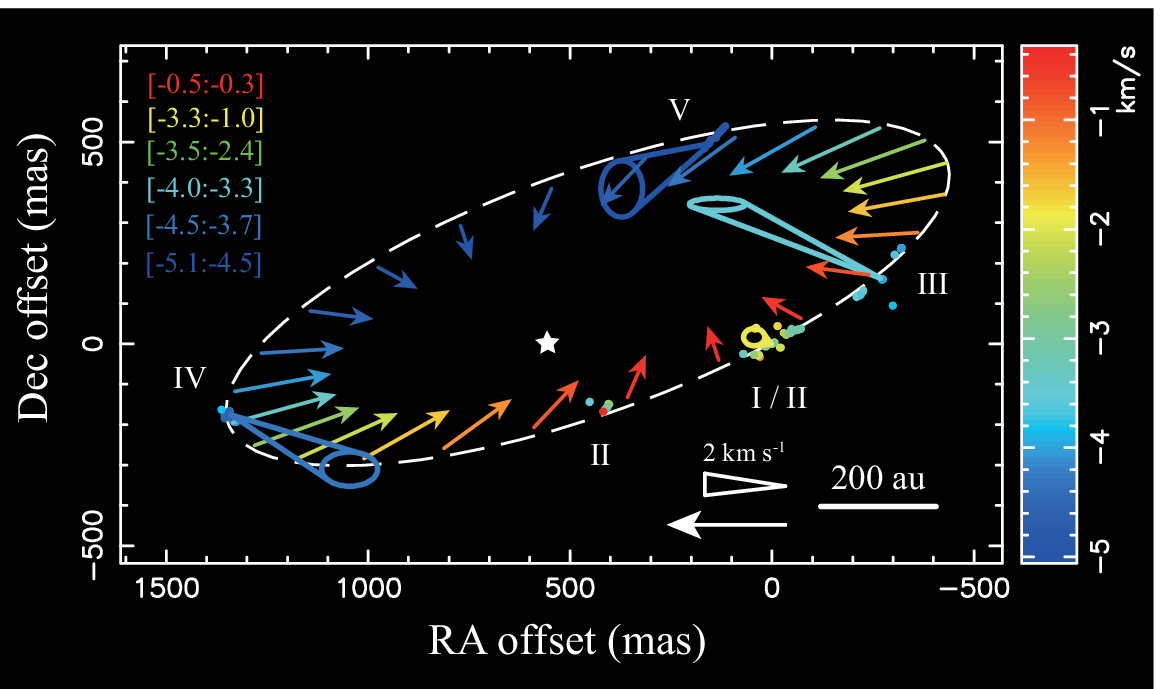}
\end{center}
\caption{{\footnotesize
Superposition plot of the maser distribution (Figure~\ref{fig2}) and the detected proper motions
(upper panel in Figure~\ref{fig5})
on the fitted disk model (lower panel in Figure~\ref{fig5}).
Colors of the maser spots and the model arrows indicate the LSR velocity
(color-velocity conversion code shown on the right of the plot).
Colors of the cones associated with 
each maser cluster indicate the LSR velocity,
with the range of the velocities within a cluster ([min:max]) shown in the top-left corner.
The scale for the amplitude of the arrows and the cones
is shown in the bottom-right corner.
}}
\label{fig6}
\end{figure*}

The lower-panel in Figure~\ref{fig5} shows the 
velocity distribution of the fitted model.
The fitted model is superposed on the maser distribution 
and the detected proper motions in Figure~\ref{fig6}.
The tangential velocity of all the maser clusters
is reproduced within the relatively large uncertainties.
Considering the l.o.s. velocity, maser clusters I/II (located near the origin), II, IV, and V 
are in good agreement with the model within the uncertainties.
The difference in the l.o.s. velocity between observation and the model for cluster III
may be caused by random motions due to turbulence driven by the magneto-rotational 
instability (see Section~\ref{section4-2} for further discussion).

The detected internal proper motions provide the information necessary to resolve the ambiguity 
in the geometry of the accretion disk and the high-velocity radio jet in Cep~A-HW2. 
Since cluster~V is located on the far side, and cluster I/II is on the near side of the accretion disk, 
the north-east component of the radio jet must be ejected in the direction of the observer 
(if it is assumed that the jet is ejected in a direction nearly perpendicular to the disk). 
This geometry is consistent with that inferred from the 3-D magnetic field structure determined 
by \citet{Vlem10}, and also with the blue- and red-shifted 
lobes of the outflow detected in the CO line emission 
(blue: north-east, red: south-west, e.g., \citealt{Rodri80}; \citealt{Cunnin09}).
The high-velocity radio jet is likely powering the low-velocity CO outflow.

We note that there is a discrepancy in the geometry derived here and that determined from 
the time delay in the synchronized flux density variations measured by \citet{Sugi08b}. 
\citeauthor{Sugi08b} observed methanol maser spectral features I-V to show 
rapid flux density variation, with the flux variation of different spectral features occurring 
with small delays. 
Time delays of $+$2, $-$1, $-$6, and $-$4~days (with respect to the strongest feature II) 
were observed for spectral features I, III, IV, and V, respectively. 
Under the assumption that the flux density variations are intrinsically synchronized 
and the time delay is caused by the light-crossing time, the observed time delay suggests 
that cluster~V is located on the near side of the disk, 
i.e., the reverse geometry to that inferred from the proper motions. 
We do not currently have a satisfactory explanation for the disagreement between 
these two independent results.
In the following, we investigate further the implications of the rotating and infalling disk 
model based on our proper motion observations.

%%% Figure 6. %%%

\subsection{Mass accretion}\label{section4-2}

The fitted rotation velocity of 0.5~km~s$^{-1}$ is less than half of the observed LSR velocity range 
($- 5.06$ to $- 0.32$~km~s$^{-1}$). 
On the other hand, the infall velocity of 1.8~km~s$^{-1}$ estimated from our model is closer 
to half of the LSR velocity range. 
This suggests that the infall motions dominate over rotation, 
at least in the vicinity of the masing zone (around a radius of $\sim$700~au).  
This implies that the Cep~A-HW2 region is in an evolutionary phase 
characterized by active gas accretion from the disk onto the high-mass YSO.

The mass infall rate can be estimated, using a method similar 
to that employed by \citealt{Goddi11}, by using the equation 
$\dot{M}_{\mathrm{inf}} = 2 \pi R T n_{\mathrm{H}_2} m_{\mathrm{H}_2} V_{\mathrm{inf}}$, 
where $R$ is the radius of the maser ring, 
$n_{\mathrm{H}_2},~m_{\mathrm{H}_2}$ are the volume density and mass of molecular hydrogen, 
$V_{\mathrm{inf}}$ is the infall velocity onto the central YSO, and $T = R \theta_{0}$.
The parameter $\theta_{0}$ is the opening angle from the central YSO to the edge of the disk. 
\citet{Vlem10} estimated the thickness of the maser disk to be $\sim$300~au, 
which implies $\theta_{0} \simeq 24 \degr$.
If we assume that $R = 680$~au and $V_{\mathrm{inf}} = 1.8$~km~s$^{-1}$, 
we obtain $\dot{M}_{\mathrm{inf}} = 3 \times 10^{-4} n_{8}~M_{\sun}~\mathrm{yr}^{-1}$, 
where $n_{8}$ is the gas volume density in units of $10^{8}~\mathrm{cm}^{-3}$. 
Observations of thermal methanol lines in Cep~A imply a gas density of 
$\sim 10^{7}~\mathrm{cm}^{-3}$ at the masing site \citep{Torst11b}. 
However, \citeauthor{Torst11b} note that the derived 
gas volume density might be diluted by the large beam ($\sim 14 \arcsec$) of their observations.
Considering the dilution factor, the true gas volume density 
could be $n_{\mathrm{H}_{2}} \geq 10^{8}~\mathrm{cm}^{-3}$
within a $2 \arcsec$ region around the Cep~A-HW2 object. 
A gas volume density of $10^{8}~\mathrm{cm}^{-3}$ is also consistent with the 
values required by the models of methanol maser excitation \citep{Cragg05}. 
Our mass infall rate is similar to that estimated by \citet{Goddi11} from their observations 
of the high-mass star forming region AFGL5142.
As indicated in the following calculations, 
this mass infall rate is high enough to continue mass accretion onto a central high-mass YSO and so this region could be forming a central star of $\sim 20~M_{\sun}$ \citep{Hosokawa09,Hosokawa10}.

Our observations and model fitting estimate the infall momentum rate 
$\dot{P}_{\mathrm{inf}} = \dot{M}_{\mathrm{inf}} V_{\mathrm{inf}}$ to be 
$5 \times 10^{-4} n_{8}~M_{\sun}~\mathrm{yr}^{-1}~\mathrm{km~s}^{-1}$. 
On the other hand, the momentum output rate produced by the stellar radiation pressure can 
be estimated using $\dot{L}_{\ast} = (L_{\ast} \tau) / (4 \pi c)$, 
where $L_{\ast}$ is the stellar luminosity, $\tau$ the average optical depth of accreting gas, 
and $c$ the velocity of light.
Taking $L_{\ast} = 1.3 \times 10^{4}~L_{\sun}$ for the Cep~A-HW2 object and $\tau = 10$ 
as an upper limit in a typical star forming region \citep{Lamers99}, the momentum output rate 
of the stellar radiation pressure is estimated to be 
$\dot{L}_{\ast} = 2 \times 10^{-4}~M_{\sun}~\mathrm{yr}^{-1}~\mathrm{km~s}^{-1}$.
Consequently, the infall momentum rate exceeds the stellar radiation pressure 
($\dot{P}_{\mathrm{inf}} \geq \dot{L}_{\ast}$), which is consistent with the infall motion 
observed at the radius of the maser emission.

The infall might be caused by the transportation of angular momentum owing to 
turbulence driven by the magneto-rotational instability \citep{Balbus91}. 
The turbulent velocity can be estimated from a comparison of the magnetic energy density 
$u_{m}$ with the kinetic energy density due to the turbulence $u_{kt}$. 
The magnetic energy density at the masing site can be estimated using the equation 
$u_{m} = B^{2} / (2 \mu_{0})$, where $B$ is the magnetic field strength 
and $\mu_{0}$ the magnetic permeability in a vacuum.
Taking $B = 8.1$~mG ($8.1 \times 10^{-7}$~T) at the masing site in Cep~A-HW2, 
as measured by \citet{Vlem08}, 
the magnetic energy density $u_{m}$ is $2.6 \times 10^{-7}~\mathrm{J~m}^{-3}$. 
In the case of $u_{m} \simeq u_{kt}$, the turbulent velocity $v_{\mathrm{turb}}$ can be 
estimated to be $\sim$1.3~km~s$^{-1}$.
The estimated turbulent velocity is comparable with the turbulent linewidth of 1.1~km~s$^{-1}$ 
calculated using the maser polarization, radiative transfer model for the methanol masers in 
Cep~A-HW2 \citep{Vlem10}.
This level of turbulence exceeds the estimated uncertainties 
for the disk model velocities (0.7~km~s$^{-1}$). 
Consequently, it is plausible that random motions could be caused by magnetically-induced 
turbulence and could account for the differences
between the model and observed maser velocities occurring 
in some parts of the system.
This turbulence could be responsible for the measured infall velocities in maser regions, 
where the fitted rotational velocity of 0.5~km~s$^{-1}$ is smaller than 
the Keplerian velocity gradient of $\sim$6~km~s$^{-1}$ derived from 
the l.o.s. velocity of the CH$_{3}$CN line emission.

The differences observed in the kinematics of the 6.7~GHz methanol maser
and CH$_{3}$CN line emissions, 
despite the very similar ring structures traced in both emissions (see Table~\ref{tab:tab4}),
might be due to their arising in different locations within the disk 
because of differing gas volume densities.
The 6.7~GHz methanol masers can arise at high densities 
($n_{\mathrm{H}_{2}} \geq 10^{8}~\mathrm{cm}^{-3}$) and may occur in infalling gas located at 
the mid-plane of the disk, where dense neutral materials are self-shielded from ionizing photons 
and can participate in the gas accretion onto the central YSO \citep[e.g.,][]{Bik08}. 
On the other hand, the CH$_{3}$CN emission is associated with moderate density gas 
($\sim 10^{6}~\mathrm{cm}^{-3}$) and may arise on the surface of the disk 
where it is strongly affected by the ionizing photon pressure. 
It is difficult for the gas in these regions to be accreted 
and hence no infall motions are expected.

\subsection{Enclosed mass}\label{section4-3}

If we assume Keplarian rotation of the maser disk then we can estimate the enclosed mass 
using the expression $M = (R V_{\mathrm{rot}}^{2}) / \mathrm{G}$, 
where G is the gravitational constant.  
For $V_{\mathrm{rot}} = 0.5~\mathrm{km~s}^{-1}$ (implied from our modeling), 
this yields an enclosed mass of $\sim 0.2~M_{\sun}$. 
An alternative method for calculating the enclosed mass uses the free fall velocity 
and the equation $M = (R V_{\mathrm{inf}}^{2}) / (2 \mathrm{G})$.  
For $V_{\mathrm{inf}} = 1.8~\mathrm{km~s}^{-1}$, the implied enclosed mass is $\sim 1.2~M_{\sun}$. 
Both of these estimates are significantly smaller than those of a high-mass star ($\geq 8 \,M_{\sun}$), 
and they are also much smaller than the enclosed mass inferred from the velocity gradient observed 
in the CH$_3$CN line emissions ($\sim 20~M_{\sun}$). 
We discuss two possible explanations as to why the masses estimated from the 3-D velocities 
of the methanol masers may be smaller than those determined from the thermal lines: 
radiation pressure and magnetic pressure. 
If the infall momentum rate is roughly equal to the momentum output rate produced 
by a stellar radiation pressure (i.e. $\dot{P}_{\mathrm{inf}} \simeq \dot{L}_{\ast}$), 
the apparent enclosed mass will approach $\sim 0~M_{\sun}$ in the free fall model. 
For Cep~A-HW2, we have shown that $\dot{P}_{\mathrm{inf}}$ is only a factor 2 
or so higher than $\dot{L}_{\ast}$ (see Section~\ref{section4-2}), and thus radiation pressure 
may play a role in supporting the cloud against infall, 
which in turn reduces the apparent enclosed mass. 
To see if magnetic pressure provides an alternative explanation, we need to compare 
the kinetic energy density (due to the rotation of the methanol gas) to the magnetic energy density. 
The kinetic energy density of the methanol gas is given by 
$u_{kr} = (1/2) n_{\mathrm{H}_2} m_{\mathrm{H}_2} V_{\mathrm{rot}}^{2}$, which equates to 
$4.2 \times 10^{-8}~\mathrm{J~m}^{-3}$ for Cep~A-HW2. 
The magnetic energy density at the masing site is estimated to be 
$2.6 \times 10^{-7}~\mathrm{J~m}^{-3}$ (see Section~\ref{section4-2}).
Since $u_{kr} \leq u_{m}$, 
the disk gas may be partially supported by the magnetic field at the radius of the maser ring 
in which case the rotation speed estimated from the methanol maser would be slower 
than that for the CH$_3$CN emission. 
This is predicted by 
the magnetic braking theory (which appears to work for low-mass star formation regions), 
according to which an efficient way of transporting angular momentum is through 
outgoing helical Alfv$\acute{\mathrm{e}}$n waves \citep{Basu94}. 
For high-mass star formation \citet{Peters11} suggest that magnetic braking might become 
qualitatively important for the local accretion onto YSOs, 
despite it being of minor importance for the global gas dynamics.
This magnetic support may be occurring only at the higher densities 
associated with the maser-gas clouds 
($n_{\mathrm{H}_{2}} \geq 10^{8}~\mathrm{cm}^{-3}$), while not being relevant 
in the moderately dense CH$_{3}$CN gas regions ($\sim 10^{6}~\mathrm{cm}^{-3}$),
if $B$ scales with the empirical relation $B \propto n_{\mathrm{H}_{2}}^{0.47}$
\citep{Crutcher99,Vlem08}.
This hypothesis can be tested in the near future through dust polarization measurement 
with the SMA to derive more accurate values of $n_{\mathrm{H}_{2}}$ and $B$ 
in the CH$_{3}$CN emission region.

%%% Section 5. Conclusions %%%
\section{Conclusions}\label{section5}

We have used three epochs of JVN observations over the period between 2006--2008 
(spanning a total of 779 days) to measure the spatial distribution and internal proper motions 
of the 6.7~GHz methanol maser spots associated with the Cep~A-HW2 YSO. 
The maser spots were found to be distributed in a curved structure surrounding the HW2 object, 
similar to previous results obtained by other authors. 
The maser distribution closely matches that of the observed CH$_3$CN and NH$_3$ disks 
in terms of the position, size, elongation, and the range of the radial velocities.

We measured internal proper motions 
in a total of 29 maser spots, and the amplitude of the proper motions ranged 
from 0.2 to 7.4~km~s$^{-1}$, with an average value of 3.1~km~s$^{-1}$.
The detected motions were modeled as originating in a 
disk with both rotation and infall velocity components. 
The derived rotation and infalling velocities at the disk radius of 680~au 
for our observations are $V_{\mathrm{rot}} = 0.5 \pm 0.7$~km~s$^{-1}$ and 
$V_{\mathrm{inf}} = 1.8 \pm 0.7$~km~s$^{-1}$, respectively.
The magnitude of the infall velocity compared to the rotational component suggests 
that the Cep~A-HW2 object is at an evolutionary phase
where there is active gas accretion from the disk onto the central high-mass YSO. 
The mass infall rate estimated from the derived infall velocity is consistent 
with the possibility of the HW2 object having a final mass of $\sim 20~M_{\sun}$, 
and the infall momentum rate appears sufficient to overcome the stellar radiation pressure.

%__________________________________________________________________

\begin{acknowledgements}
The authors thank the JVN team for assistance and support during the observations.
The authors also thank the anonymous referee
for many useful suggestions and comments, which improved this paper.
The JVN project is led by the National Astronomical Observatory
of Japan, which is a branch of the National Institutes of
Natural Sciences, Hokkaido University, The University of Tsukuba,
Ibaraki University, Gifu University, Osaka Prefecture University,
Yamaguchi University, and Kagoshima University,
in cooperation with the Geospatial Information Authority of Japan, 
the Japan Aerospace Exploration Agency, and the National
Institute of Information and Communications Technology. 
\end{acknowledgements}

\bibliographystyle{aa}
\bibliography{sugiyamabib}

%%% Table 5. %%%
%%% available electronically only %%%
\onltab{5}{
\onecolumn
\begin{landscape}
\begin{longtable}{rrrrrrrrrrrrrrrrr}
\caption{Parameters of the 6.7~GHz methanol maser spots in Cep~A.
The relative positional offset in columns 3 and 5 is obtained with respect to
the reference spot (ID 53 marked by an asterisk), whose absolute coordinates are 
[$\alpha (\mathrm{J2000.0}) = 22^{\mathrm{h}} 56^{\mathrm{m}} 17^{\mathrm{s}}.9042$, $\delta (\mathrm{J2000.0}) = +62 \degr 01 \arcmin49{\arcsec}.577$].
Proper motions are measured relative to the maser barycenter 
(see Section~\ref{section3-2}).
}\label{longtab1}
\\
\hline \hline
\multicolumn{1}{c}{ID}
& \multicolumn{1}{c}{Cluster}
& \multicolumn{4}{c}{Relative positional offset}
& \multicolumn{1}{c}{LSR}
& \multicolumn{6}{c}{Internal proper motion}
&
& \multicolumn{3}{c}{Peak intensity}
\\ \cline{3-6} \cline{8-13} \cline{15-17}
&
& \multicolumn{1}{c}{RA}
& \multicolumn{1}{c}{$\sigma_{\alpha}$}
& \multicolumn{1}{c}{Dec}
& \multicolumn{1}{c}{$\sigma_{\delta}$}
& \multicolumn{1}{c}{Velocity}
& \multicolumn{1}{c}{$\mu_{\alpha}$}
& \multicolumn{1}{c}{$\sigma \mu_{\alpha}$}
& \multicolumn{1}{c}{$V_{x}$}
& \multicolumn{1}{c}{$\mu_{\delta}$}
& \multicolumn{1}{c}{$\sigma \mu_{\delta}$}
& \multicolumn{1}{c}{$V_{y}$}
&
& \multicolumn{1}{c}{1st}
& \multicolumn{1}{c}{2nd}
& \multicolumn{1}{c}{3rd}
\\ \cline{3-6} \cline{8-9} \cline{11-12} \cline{15-17}
&
& \multicolumn{4}{c}{(mas)}
& \multicolumn{1}{c}{(km~s$^{-1}$)}
& \multicolumn{2}{c}{(mas~yr$^{-1}$)}
& \multicolumn{1}{c}{(km~s$^{-1}$)}
& \multicolumn{2}{c}{(mas~yr$^{-1}$)}
& \multicolumn{1}{c}{(km~s$^{-1}$)}
&
& \multicolumn{3}{c}{(Jy~beam$^{-1}$)}
\\ \hline
\endfirsthead
\caption{Continued.}\\
\hline\hline
\multicolumn{1}{c}{ID}
& \multicolumn{1}{c}{Cluster}
& \multicolumn{4}{c}{Relative positional offset}
& \multicolumn{1}{c}{LSR}
& \multicolumn{6}{c}{Internal proper motion}
&
& \multicolumn{3}{c}{Peak intensity}
\\ \cline{3-6} \cline{8-13} \cline{15-17}
&
& \multicolumn{1}{c}{RA}
& \multicolumn{1}{c}{$\sigma_{\alpha}$}
& \multicolumn{1}{c}{Dec}
& \multicolumn{1}{c}{$\sigma_{\delta}$}
& \multicolumn{1}{c}{Velocity}
& \multicolumn{1}{c}{$\mu_{\alpha}$}
& \multicolumn{1}{c}{$\sigma \mu_{\alpha}$}
& \multicolumn{1}{c}{$V_{x}$}
& \multicolumn{1}{c}{$\mu_{\delta}$}
& \multicolumn{1}{c}{$\sigma \mu_{\delta}$}
& \multicolumn{1}{c}{$V_{y}$}
&
& \multicolumn{1}{c}{1st}
& \multicolumn{1}{c}{2nd}
& \multicolumn{1}{c}{3rd}
\\ \cline{3-6} \cline{8-9} \cline{11-12} \cline{15-17}
&
& \multicolumn{4}{c}{(mas)}
& \multicolumn{1}{c}{(km~s$^{-1}$)}
& \multicolumn{2}{c}{(mas~yr$^{-1}$)}
& \multicolumn{1}{c}{(km~s$^{-1}$)}
& \multicolumn{2}{c}{(mas~yr$^{-1}$)}
& \multicolumn{1}{c}{(km~s$^{-1}$)}
&
& \multicolumn{3}{c}{(Jy~beam$^{-1}$)}
\\ \hline
\endhead
\hline
\endfoot
\hline
\endlastfoot
\footnotesize
1 & III & $-$320.49  & 0.68  & 237.24  & 0.46  & $-$3.83   & $\cdots$ & $\cdots$ & $\cdots$ & $\cdots$ & $\cdots$ & $\cdots$  &  & 1.45  & 7.16  & $\cdots$ \\
2 & III & $-$319.99  & 0.41  & 236.96  & 0.31  & $-$3.65    & $\cdots$ & $\cdots$ & $\cdots$ & $\cdots$ & $\cdots$ & $\cdots$  &  & 2.10  & 7.51  & $\cdots$ \\
3 & III & $-$319.45  & 0.45  & 236.78  & 0.34  & $-$3.48    & $\cdots$ & $\cdots$ & $\cdots$ & $\cdots$ & $\cdots$ & $\cdots$  &  & 1.40  & 3.55  & $\cdots$ \\
4 & III & $-$317.90  & 0.59  & 234.75  & 0.49  & $-$4.01    & $\cdots$ & $\cdots$ & $\cdots$ & $\cdots$ & $\cdots$ & $\cdots$ &  & $\cdots$ & 2.72  & $\cdots$ \\
5 & III & $-$303.73  & 0.40  & 220.89  & 0.32  & $-$3.48    & $\cdots$ & $\cdots$ & $\cdots$ & $\cdots$ & $\cdots$ & $\cdots$ &  & 1.33  & $\cdots$ & $\cdots$ \\
6 & III & $-$303.30  & 0.30  & 220.80  & 0.19  & $-$3.65    & $\cdots$ & $\cdots$ & $\cdots$ & $\cdots$ & $\cdots$ & $\cdots$ &  & 2.13  & $\cdots$ & $\cdots$ \\
7 & III & $-$299.02  & 0.74  & 94.65  & 0.30  & $-$3.83    & $\cdots$ & $\cdots$ & $\cdots$ & $\cdots$ & $\cdots$ & $\cdots$ &  & $\cdots$ & $\cdots$ & 3.50  \\
8 & III & $-$298.73  & 0.86  & 94.70  & 0.34  & $-$3.65    & $\cdots$ & $\cdots$ & $\cdots$ & $\cdots$ & $\cdots$ & $\cdots$ &  & $\cdots$ & $\cdots$ & 3.53  \\
9 & III & $-$274.34  & 0.25  & 160.71  & 0.13  & $-$3.48      & 1.57 & 0.40 & 5.21 & 0.46 & 0.19 & 1.52  &  & 2.64  & 1.90  & 2.42  \\
10 & III & $-$273.89  & 0.14  & 160.43  & 0.10  & $-$3.65    & $\cdots$ & $\cdots$ & $\cdots$ & $\cdots$ & $\cdots$ & $\cdots$  &  & 5.60  & 6.30  & 8.32  \\
11 & III & $-$273.22  & 0.26  & 159.77  & 0.11  & $-$3.83    & 0.98 & 0.19 & 3.26 & 0.54 & 0.08 & 1.80  &  & 4.79  & 8.40  & 10.41  \\
12 & III & $-$271.77  & 0.51  & 158.62  & 0.27  & $-$4.01    & 1.60 & 0.30 & 5.31 & 0.69 & 0.18 & 2.29  &  & 2.30  & 3.83  & 8.02  \\
13 & III & $-$225.53  & 0.51  & 132.59  & 0.52  & $-$3.30    & $\cdots$ & $\cdots$ & $\cdots$ & $\cdots$ & $\cdots$ & $\cdots$ &  & $\cdots$ & 1.01  & $\cdots$ \\
14 & III & $-$224.40  & 0.40  & 129.49  & 0.45  & $-$3.48    & $\cdots$ & $\cdots$ & $\cdots$ & $\cdots$ & $\cdots$ & $\cdots$ &  & $\cdots$ & 1.88  & $\cdots$ \\
15 & III & $-$219.96  & 0.21  & 121.00  & 0.26  & $-$3.65    & $\cdots$ & $\cdots$ & $\cdots$ & $\cdots$ & $\cdots$ & $\cdots$ &  & $\cdots$ & 4.96  & $\cdots$ \\
16 & III & $-$216.10  & 0.27  & 119.06  & 0.16  & $-$3.48    & $\cdots$ & $\cdots$ & $\cdots$ & $\cdots$ & $\cdots$ & $\cdots$  &  & 2.21  & 3.83  & $\cdots$ \\
17 & III & $-$214.45  & 0.23  & 119.65  & 0.15  & $-$3.30    & $\cdots$ & $\cdots$ & $\cdots$ & $\cdots$ & $\cdots$ & $\cdots$  &  & 1.46  & 1.53  & $\cdots$ \\
18 & III & $-$209.55  & 0.33  & 117.36  & 0.33  & $-$3.30    & $\cdots$ & $\cdots$ & $\cdots$ & $\cdots$ & $\cdots$ & $\cdots$ &  & $\cdots$ & 1.43  & $\cdots$ \\
19 & III & $-$208.87  & 0.21  & 115.73  & 0.22  & $-$3.48    & $\cdots$ & $\cdots$ & $\cdots$ & $\cdots$ & $\cdots$ & $\cdots$ &  & $\cdots$ & 2.59  & $\cdots$ \\
20 & I/II & $-$72.49  & 0.27  & 38.53  & 0.14  & $-$3.13    & $\cdots$ & $\cdots$ & $\cdots$ & $\cdots$ & $\cdots$ & $\cdots$  &  & 2.08  & 3.74  & 3.18  \\
21 & I/II & $-$71.69  & 0.11  & 37.99  & 0.05  & $-$2.95    & $\cdots$ & $\cdots$ & $\cdots$ & $\cdots$ & $\cdots$ & $\cdots$  &  & 10.04  & 11.73  & 9.71  \\
22 & I/II & $-$70.24  & 0.13  & 37.35  & 0.07  & $-$2.78    & $\cdots$ & $\cdots$ & $\cdots$ & $\cdots$ & $\cdots$ & $\cdots$  &  & 16.92  & 13.94  & 8.59  \\
23 & I/II & $-$68.47  & 0.12  & 36.38  & 0.06  & $-$2.60    & $\cdots$ & $\cdots$ & $\cdots$ & $\cdots$ & $\cdots$ & $\cdots$  &  & 13.36  & $\cdots$ & 11.69  \\
24 & I/II & $-$67.42  & 0.14  & 37.00  & 0.09  & $-$2.95    & $\cdots$ & $\cdots$ & $\cdots$ & $\cdots$ & $\cdots$ & $\cdots$  &  & $\cdots$ & 12.22  & 8.40  \\
25 & I/II & $-$65.30  & 0.19  & 35.43  & 0.15  & $-$2.78    & $\cdots$ & $\cdots$ & $\cdots$ & $\cdots$ & $\cdots$ & $\cdots$  &  & $\cdots$ & 19.11  & 13.72  \\
26 & I/II & $-$61.87  & 0.17  & 35.37  & 0.09  & $-$2.95    & $\cdots$ & $\cdots$ & $\cdots$ & $\cdots$ & $\cdots$ & $\cdots$  &  & 6.97  & 14.15  & 4.93  \\
27 & I/II & $-$60.54  & 0.12  & 34.43  & 0.06  & $-$2.78    & $\cdots$ & $\cdots$ & $\cdots$ & $\cdots$ & $\cdots$ & $\cdots$  &  & 19.02  & 31.03  & 9.13  \\
28 & I/II & $-$60.41  & 0.50  & 34.83  & 0.28  & $-$3.13    & $\cdots$ & $\cdots$ & $\cdots$ & $\cdots$ & $\cdots$ & $\cdots$  &  & 1.12  & 2.74  & $\cdots$ \\
29 & I/II & $-$59.37  & 0.08  & 33.63  & 0.03  & $-$2.60    & $\cdots$ & $\cdots$ & $\cdots$ & $\cdots$ & $\cdots$ & $\cdots$  &  & 22.60  & 26.64  & $\cdots$ \\
30 & I/II & $-$59.25  & 0.26  & 33.52  & 0.11  & $-$2.42    & $\cdots$ & $\cdots$ & $\cdots$ & $\cdots$ & $\cdots$ & $\cdots$ &  & 9.87  & $\cdots$ & $\cdots$ \\
31 & I/II & $-$48.20  & 0.15  & 28.78  & 0.14  & $-$2.95    & $\cdots$ & $\cdots$ & $\cdots$ & $\cdots$ & $\cdots$ & $\cdots$ &  & $\cdots$ & 7.55  & 6.29  \\
32 & I/II & $-$47.33  & 0.48  & 37.66  & 0.39  & $-$3.13    & $\cdots$ & $\cdots$ & $\cdots$ & $\cdots$ & $\cdots$ & $\cdots$ &  & $\cdots$ & $\cdots$ & 1.11  \\
33 & I/II & $-$46.39  & 0.15  & 28.04  & 0.08  & $-$2.78    & $\cdots$ & $\cdots$ & $\cdots$ & $\cdots$ & $\cdots$ & $\cdots$  &  & 13.59  & 25.88  & 18.06  \\
34 & I/II & $-$45.23  & 0.06  & 27.48  & 0.03  & $-$2.60    & $\cdots$ & $\cdots$ & $\cdots$ & $\cdots$ & $\cdots$ & $\cdots$  &  & 24.24  & 34.45  & 25.46  \\
35 & I/II & $-$44.96  & 0.17  & 27.57  & 0.09  & $-$2.42    & $\cdots$ & $\cdots$ & $\cdots$ & $\cdots$ & $\cdots$ & $\cdots$  &  & 17.14  & 12.80  & 14.29  \\
36 & I/II & $-$42.18  & 0.16  & 26.38  & 0.12  & $-$2.78    & $\cdots$ & $\cdots$ & $\cdots$ & $\cdots$ & $\cdots$ & $\cdots$  &  & $\cdots$ & 19.93  & 15.09  \\
37 & I/II & $-$40.79  & 0.09  & 25.79  & 0.07  & $-$2.60    & $\cdots$ & $\cdots$ & $\cdots$ & $\cdots$ & $\cdots$ & $\cdots$  &  & $\cdots$ & 35.11  & 25.68  \\
38 & I/II & $-$40.37  & 0.18  & 25.80  & 0.09  & $-$2.42    & $\cdots$ & $\cdots$ & $\cdots$ & $\cdots$ & $\cdots$ & $\cdots$  &  & 14.22  & 21.52  & 18.62  \\
39 & I/II & $-$34.91  & 0.20  & 23.20  & 0.16  & $-$2.42    & $\cdots$ & $\cdots$ & $\cdots$ & $\cdots$ & $\cdots$ & $\cdots$ &  & $\cdots$ & 17.96  & $\cdots$ \\
40 & I/II & $-$34.73  & 0.18  & 22.63  & 0.19  & $-$2.25    & $\cdots$ & $\cdots$ & $\cdots$ & $\cdots$ & $\cdots$ & $\cdots$ &  & $\cdots$ & 15.54  & 9.31  \\
41 & I/II & $-$28.73  & 0.29  & 28.04  & 0.16  & $-$2.25    & $\cdots$ & $\cdots$ & $\cdots$ & $\cdots$ & $\cdots$ & $\cdots$ &  & $\cdots$ & $\cdots$ & 6.65  \\
42 & I/II & $-$21.21  & 0.19  & $-$8.92  & 0.10  & $-$2.07  & $\cdots$ & $\cdots$ & $\cdots$ & $\cdots$ & $\cdots$ & $\cdots$ &  & 10.22  & $\cdots$ & $\cdots$ \\
43 & I/II & $-$19.97  & 0.23  & $-$9.64  & 0.13  & $-$1.90   & $\cdots$ & $\cdots$ & $\cdots$ & $\cdots$ & $\cdots$ & $\cdots$ &  & 7.57  & $\cdots$ & $\cdots$ \\
44 & I/II & $-$13.36  & 0.45  & 44.29  & 0.43  & $-$1.72    & $\cdots$ & $\cdots$ & $\cdots$ & $\cdots$ & $\cdots$ & $\cdots$ &  & $\cdots$ & 2.78  & $\cdots$ \\
45 & I/II & $-$13.02  & 0.51  & 44.62  & 0.49  & $-$1.55    & $\cdots$ & $\cdots$ & $\cdots$ & $\cdots$ & $\cdots$ & $\cdots$ &  & $\cdots$ & 1.18  & $\cdots$ \\
46 & I/II & $-$6.27    & 0.07  & 3.04    & 0.04   & $-$2.07    & 0.33 & 0.09 & 1.09 & 0.16 & 0.07 & 0.54  &  & 24.59  & 21.44  & 15.12  \\
47 & I/II & $-$5.41  & 0.05  & 2.56  & 0.03  & $-$2.25         & 0.35 & 0.04 & 1.15 & 0.34 & 0.03 & 1.12  &  & 46.03  & 48.59  & 35.85  \\
48 & I/II & $-$4.74  & 0.04  & 2.90  & 0.03  & $-$2.60    & $\cdots$ & $\cdots$ & $\cdots$ & $\cdots$ & $\cdots$ & $\cdots$ &  & $\cdots$ & $\cdots$ & 27.33  \\
49 & I/II & $-$4.10  & 0.19  & 2.39  & 0.11  & $-$1.90    & $\cdots$ & $\cdots$ & $\cdots$ & $\cdots$ & $\cdots$ & $\cdots$ &  & 8.70  & $\cdots$ & $\cdots$ \\
50 & I/II & $-$2.04  & 0.04  & 0.91  & 0.02  & $-$2.42    & $-$0.52 & 0.05 & $-$1.74 & 0.59 & 0.03 & 1.97  &  & 63.74  & 73.17  & 45.07  \\
51 & I/II & $-$0.94  & 0.46  & 0.31  & 0.24  & $-$3.30    & $\cdots$ & $\cdots$ & $\cdots$ & $\cdots$ & $\cdots$ & $\cdots$  &  & 0.85  & 0.93  & $\cdots$ \\
52 & I/II & $-$0.57  & 0.33  & 0.68  & 0.17  & $-$3.13    & 1.35 & 0.24 & 4.47 & $-$0.45 & 0.13 & $-$1.49  &  & 1.75  & 2.25  & 1.53  \\
53$^{\ast}$ & I/II & 0.00  & 0.02  & 0.00  & 0.01  & $-$2.60    & 0.13 & 0.02 & 0.44 & $-$0.03 & 0.01 & $-$0.10   &  & 68.78  & 92.19  & 46.06  \\
54 & I/II & 0.88  & 0.15  & 0.07  & 0.07  & $-$2.95    & $\cdots$ & $\cdots$ & $\cdots$ & $\cdots$ & $\cdots$ & $\cdots$  &  & 7.61  & 13.78  & 8.82  \\
55 & I/II & 0.99  & 0.07  & $-$0.29  & 0.03  & $-$2.78    & 0.06 & 0.04 & 0.20 & $-$0.02 & 0.02 & $-$0.07  &  & 36.96  & 55.84  & 29.14  \\
56 & I/II & 15.43  & 0.47  & $-$5.97  & 0.27  & $-$2.78    & $\cdots$ & $\cdots$ & $\cdots$ & $\cdots$ & $\cdots$ & $\cdots$ &  & 5.23  & $\cdots$ & $\cdots$ \\
57 & I/II & 16.35  & 0.25  & $-$6.69  & 0.14  & $-$2.60    & $\cdots$ & $\cdots$ & $\cdots$ & $\cdots$ & $\cdots$ & $\cdots$ &  & 6.12  & $\cdots$ & $\cdots$ \\
58 & I/II & 29.20  & 0.75  & $-$31.63  & 0.46  & $-$1.19    & $\cdots$ & $\cdots$ & $\cdots$ & $\cdots$ & $\cdots$ & $\cdots$ &  & 0.43  & $\cdots$ & $\cdots$ \\
59 & I/II & 32.44  & 0.86  & $-$28.53  & 0.49  & $-$1.02    & $\cdots$ & $\cdots$ & $\cdots$ & $\cdots$ & $\cdots$ & $\cdots$ &  & 0.37  & $\cdots$ & $\cdots$ \\
60 & I/II & 32.77  & 0.24  & $-$28.11  & 0.14  & $-$2.07    & $\cdots$ & $\cdots$ & $\cdots$ & $\cdots$ & $\cdots$ & $\cdots$  &  & 6.79  & 8.58  & 8.96  \\
61 & I/II & 32.93  & 0.07  & $-$28.36  & 0.04  & $-$1.90    & $\cdots$ & $\cdots$ & $\cdots$ & $\cdots$ & $\cdots$ & $\cdots$  &  & 25.63  & 28.96  & 18.07  \\
62 & I/II & 33.61  & 0.07  & $-$28.61  & 0.03  & $-$1.72    & $\cdots$ & $\cdots$ & $\cdots$ & $\cdots$ & $\cdots$ & $\cdots$  &  & 34.02  & 33.90  & 17.00  \\
63 & I/II & 33.62  & 0.37  & $-$27.30  & 0.21  & $-$1.19    & $\cdots$ & $\cdots$ & $\cdots$ & $\cdots$ & $\cdots$ & $\cdots$ &  & 0.87  & $\cdots$ & $\cdots$ \\
64 & I/II & 34.14  & 0.08  & $-$28.72  & 0.04  & $-$1.55    & $\cdots$ & $\cdots$ & $\cdots$ & $\cdots$ & $\cdots$ & $\cdots$  &  & 18.96  & 13.93  & 5.97  \\
65 & I/II & 34.95  & 0.12  & $-$28.73  & 0.05  & $-$1.37    & $-$0.03 & 0.21 & $-$0.11 & 0.30 & 0.13 & 1.01   &  & 3.53  & 1.41  & 0.75  \\
66 & I/II & 39.13  & 0.63  & 39.68  & 0.50  & $-$1.90    & $\cdots$ & $\cdots$ & $\cdots$ & $\cdots$ & $\cdots$ & $\cdots$ &  & $\cdots$ & 5.15  & $\cdots$ \\
67 & I/II & 40.76  & 0.43  & 39.13  & 0.41  & $-$1.72    & $\cdots$ & $\cdots$ & $\cdots$ & $\cdots$ & $\cdots$ & $\cdots$ &  & $\cdots$ & 3.38  & $\cdots$ \\
68 & I/II & 41.41  & 0.41  & 38.33  & 0.41  & $-$1.55    & $\cdots$ & $\cdots$ & $\cdots$ & $\cdots$ & $\cdots$ & $\cdots$ &  & $\cdots$ & 1.64  & $\cdots$ \\
69 & I/II & 41.60  & 0.32  & $-$25.99  & 0.19  & $-$2.07    & $\cdots$ & $\cdots$ & $\cdots$ & $\cdots$ & $\cdots$ & $\cdots$ &  & 5.88  & $\cdots$ & $\cdots$ \\
70 & I/II & 41.71  & 0.21  & $-$26.99  & 0.13  & $-$1.90    & $\cdots$ & $\cdots$ & $\cdots$ & $\cdots$ & $\cdots$ & $\cdots$ &  & 8.44  & $\cdots$ & $\cdots$ \\
71 & I/II & 42.39  & 0.55  & 37.45  & 0.45  & $-$1.37    & $\cdots$ & $\cdots$ & $\cdots$ & $\cdots$ & $\cdots$ & $\cdots$ &  & $\cdots$ & 0.63  & $\cdots$ \\
72 & I/II & 43.88  & 0.41  & $-$27.14  & 0.31  & $-$2.60    & $\cdots$ & $\cdots$ & $\cdots$ & $\cdots$ & $\cdots$ & $\cdots$ &  & $\cdots$ & 7.10  & $\cdots$ \\
73 & I/II & 44.57  & 0.58  & $-$27.80  & 0.43  & $-$2.42    & $\cdots$ & $\cdots$ & $\cdots$ & $\cdots$ & $\cdots$ & $\cdots$ &  & $\cdots$ & 5.81  & $\cdots$ \\
74 & I/II & 70.86  & 0.28  & $-$24.90  & 0.21  & $-$2.95    & $\cdots$ & $\cdots$ & $\cdots$ & $\cdots$ & $\cdots$ & $\cdots$ &  & 3.33  & 3.28  & $\cdots$ \\
75 & I/II & 71.10  & 0.28  & $-$24.70  & 0.18  & $-$3.30    & 0.13 & 0.30 & 0.42 & 0.86 & 0.14 & 2.86   &  & 1.26  & 2.74  & 1.33  \\
76 & I/II & 71.25  & 0.15  & $-$24.85  & 0.10  & $-$3.13    & 0.44 & 0.11 & 1.47 & 0.46 & 0.07 & 1.53   &  & 3.13  & 5.32  & 2.38  \\
77 & V & 115.92  & 0.48  & 538.56  & 0.41  & $-$4.88    & $\cdots$ & $\cdots$ & $\cdots$ & $\cdots$ & $\cdots$ & $\cdots$ &  & $\cdots$ & $\cdots$ & 1.09  \\
78 & V & 120.52  & 0.67  & 534.02  & 0.71  & $-$5.06    & $\cdots$ & $\cdots$ & $\cdots$ & $\cdots$ & $\cdots$ & $\cdots$ &  & $\cdots$ & 0.59  & $\cdots$ \\
79 & V & 121.12  & 0.26  & 533.02  & 0.27  & $-$4.88    & 0.62 & 0.24 & 2.04 & 0.09 & 0.24 & 0.29     &  & $\cdots$ & 1.68  & 2.93  \\
80 & V & 122.16  & 0.34  & 532.20  & 0.35  & $-$4.71    & 0.36 & 0.66 & 1.20 & 0.41 & 0.69 & 1.34     &  & $\cdots$ & 1.75  & 3.29  \\
81 & V & 124.18  & 0.21  & 530.14  & 0.22  & $-$4.88    & $\cdots$ & $\cdots$ & $\cdots$ & $\cdots$ & $\cdots$ & $\cdots$ &  & $\cdots$ & $\cdots$ & 2.38  \\
82 & V & 128.07  & 0.33  & 525.42  & 0.34  & $-$4.88    & 0.53 & 0.32 & 1.74 & $-$1.03 & 0.31 & $-$3.41  &  & $\cdots$ & 1.41  & 2.15  \\
83 & V & 129.54  & 0.31  & 523.64  & 0.33  & $-$4.71    & 1.06 & 0.75 & 3.51 & 0.22 & 0.78 & 0.74  &  & $\cdots$ & 1.91  & 2.85  \\
84 & V & 133.56  & 0.40  & 520.15  & 0.31  & $-$4.88    & $\cdots$ & $\cdots$ & $\cdots$ & $\cdots$ & $\cdots$ & $\cdots$ &  & $\cdots$ & $\cdots$ & 1.32  \\
85 & V & 138.41  & 0.63  & 511.83  & 0.59  & $-$4.88    & $\cdots$ & $\cdots$ & $\cdots$ & $\cdots$ & $\cdots$ & $\cdots$ &  & $\cdots$ & 0.69  & $\cdots$ \\
86 & V & 140.27  & 0.52  & 511.04  & 0.53  & $-$4.71    & $\cdots$ & $\cdots$ & $\cdots$ & $\cdots$ & $\cdots$ & $\cdots$ &  & $\cdots$ & 1.00  & $\cdots$ \\
87 & V & 151.31  & 0.25  & 504.58  & 0.18  & $-$4.71    & 0.04 & 0.34 & 0.13 & 0.13 & 0.29 & 0.43   &  & 1.43  & 3.57  & $\cdots$ \\
88 & V & 151.40  & 0.92  & 504.26  & 0.68  & $-$4.88    & $-$0.53 & 1.08 & $-$1.75 & 0.88 & 0.84 & 2.93   &  & 0.45  & 1.56  & $\cdots$ \\
89 & V & 151.96  & 0.23  & 504.27  & 0.18  & $-$4.53    & 0.57 & 0.45 & 1.88 & $-$0.93 & 0.46 & $-$3.10   &  & 1.55  & 2.75  & $\cdots$ \\
90 & V & 157.68  & 0.23  & 496.85  & 0.19  & $-$4.53    & 1.69 & 0.46 & 5.60 & $-$1.48 & 0.49 & $-$4.89   &  & 1.51  & 2.81  & $\cdots$ \\
91 & V & 157.90  & 0.72  & 495.60  & 0.53  & $-$4.88    & 1.46 & 0.86 & 4.85 & 0.66 & 0.68 & 2.17   &  & 0.46  & 1.59  & $\cdots$ \\
92 & V & 158.16  & 0.26  & 496.37  & 0.19  & $-$4.71    & 1.11 & 0.37 & 3.69 & $-$0.83 & 0.32 & $-$2.76  &  & 1.33  & 3.10  & $\cdots$ \\
93 & V & 165.18  & 0.55  & 488.11  & 0.41  & $-$4.71    & $\cdots$ & $\cdots$ & $\cdots$ & $\cdots$ & $\cdots$ & $\cdots$  &  & 0.65  & 1.15  & $\cdots$ \\
94 & V & 167.38  & 0.69  & 483.48  & 0.81  & $-$4.88    & $\cdots$ & $\cdots$ & $\cdots$ & $\cdots$ & $\cdots$ & $\cdots$ &  & $\cdots$ & 0.58  & $\cdots$ \\
95 & II & 402.69  & 0.60  & $-$149.27  & 0.34  & $-$2.60    & $\cdots$ & $\cdots$ & $\cdots$ & $\cdots$ & $\cdots$ & $\cdots$ &  & 1.87  & $\cdots$ & $\cdots$ \\
96 & II & 405.57  & 0.82  & $-$148.12  & 0.36  & $-$2.42    & $\cdots$ & $\cdots$ & $\cdots$ & $\cdots$ & $\cdots$ & $\cdots$ &  & 2.81  & $\cdots$ & $\cdots$ \\
97 & II & 406.40  & 0.64  & $-$155.93  & 0.37  & $-$2.78    & $\cdots$ & $\cdots$ & $\cdots$ & $\cdots$ & $\cdots$ & $\cdots$ &  & 3.25  & $\cdots$ & $\cdots$ \\
98 & II & 407.92  & 0.30  & $-$156.50  & 0.18  & $-$2.95    & $\cdots$ & $\cdots$ & $\cdots$ & $\cdots$ & $\cdots$ & $\cdots$  &  & 3.44  & 1.92  & $\cdots$ \\
99 & II & 408.88  & 0.37  & $-$156.78  & 0.25  & $-$3.13    & $\cdots$ & $\cdots$ & $\cdots$ & $\cdots$ & $\cdots$ & $\cdots$ &  & 1.36  & $\cdots$ & $\cdots$ \\
100 & II & 413.36  & 1.04  & $-$161.94  & 0.71  & $-$2.78    & $\cdots$ & $\cdots$ & $\cdots$ & $\cdots$ & $\cdots$ & $\cdots$ &  & 2.08  & $\cdots$ & $\cdots$ \\
101 & II & 413.61  & 0.39  & $-$162.63  & 0.27  & $-$2.95    & $\cdots$ & $\cdots$ & $\cdots$ & $\cdots$ & $\cdots$ & $\cdots$ &  & 2.53  & $\cdots$ & $\cdots$ \\
102 & II & 413.71  & 0.45  & $-$163.00  & 0.29  & $-$3.13    & $\cdots$ & $\cdots$ & $\cdots$ & $\cdots$ & $\cdots$ & $\cdots$ &  & 1.15  & 1.11  & $\cdots$ \\
103 & II & 417.61  & 0.76  & $-$167.09  & 0.35  & $-$0.49    & $\cdots$ & $\cdots$ & $\cdots$ & $\cdots$ & $\cdots$ & $\cdots$ &  & 0.39  & $\cdots$ & $\cdots$ \\
104 & II & 418.01  & 0.50  & $-$166.97  & 0.27  & $-$0.32    & $\cdots$ & $\cdots$ & $\cdots$ & $\cdots$ & $\cdots$ & $\cdots$ &  & 0.43  & $\cdots$ & $\cdots$ \\
105 & II  & 451.96  & 0.48  & $-$142.77  & 0.54  & $-$3.48    & $\cdots$ & $\cdots$ & $\cdots$ & $\cdots$ & $\cdots$ & $\cdots$ &  & $\cdots$ & 1.06  & $\cdots$ \\
106 & II  & 452.28  & 0.49  & $-$143.34  & 0.60  & $-$3.30    & $\cdots$ & $\cdots$ & $\cdots$ & $\cdots$ & $\cdots$ & $\cdots$ &  & $\cdots$ & 0.73  & $\cdots$ \\
107 & IV & 1327.92  & 0.49  & $-$193.17  & 0.47  & $-$4.01  & $\cdots$ & $\cdots$ & $\cdots$ & $\cdots$ & $\cdots$ & $\cdots$ &  & $\cdots$ & $\cdots$ & 6.27  \\
108 & IV & 1331.99  & 0.23  & $-$193.64  & 0.15  & $-$4.18  & $\cdots$ & $\cdots$ & $\cdots$ & $\cdots$ & $\cdots$ & $\cdots$ &  & $\cdots$ & $\cdots$ & 9.09  \\
109 & IV & 1338.30  & 0.11  & $-$183.16  & 0.06  & $-$3.83  & $\cdots$ & $\cdots$ & $\cdots$ & $\cdots$ & $\cdots$ & $\cdots$  &  & 8.25  & $\cdots$ & 3.47  \\
110 & IV & 1339.17  & 0.11  & $-$183.07  & 0.06  & $-$4.01  & 0.05 & 0.13 & 0.17 & $-$0.53 & 0.07 & $-$1.76    &  & 17.80  & 5.81  & 13.41  \\
111 & IV & 1339.22  & 0.24  & $-$182.58  & 0.14  & $-$4.36  & $\cdots$ & $\cdots$ & $\cdots$ & $\cdots$ & $\cdots$ & $\cdots$  &  & 7.09  & 8.61  & $\cdots$ \\
112 & IV & 1339.60  & 0.35  & $-$182.07  & 0.20  & $-$4.53  & $\cdots$ & $\cdots$ & $\cdots$ & $\cdots$ & $\cdots$ & $\cdots$ &  & 1.28  & $\cdots$ & $\cdots$ \\
113 & IV & 1340.21  & 0.12  & $-$182.46  & 0.07  & $-$4.18  & $-$1.48 & 0.09 & $-$4.91 & $-$0.46 & 0.04 & $-$1.54    &  & 18.66  & 13.59  & 17.48  \\
114 & IV & 1343.04  & 0.16  & $-$166.12  & 0.17  & $-$4.01  & $\cdots$ & $\cdots$ & $\cdots$ & $\cdots$ & $\cdots$ & $\cdots$  &  & $\cdots$ & 6.80  & 11.68  \\
115 & IV & 1343.92  & 0.12  & $-$167.71  & 0.12  & $-$4.18  & $\cdots$ & $\cdots$ & $\cdots$ & $\cdots$ & $\cdots$ & $\cdots$ &  & $\cdots$ & 11.82  & $\cdots$ \\
116 & IV & 1344.40  & 0.11  & $-$168.19  & 0.13  & $-$4.36  & $\cdots$ & $\cdots$ & $\cdots$ & $\cdots$ & $\cdots$ & $\cdots$ &  & $\cdots$ & 8.10  & 9.14  \\
117 & IV & 1344.84  & 0.18  & $-$168.65  & 0.22  & $-$4.53  & $\cdots$ & $\cdots$ & $\cdots$ & $\cdots$ & $\cdots$ & $\cdots$ &  & $\cdots$ & 2.76  & 2.61  \\
118 & IV & 1344.95  & 0.11  & $-$179.55  & 0.06  & $-$4.36   & $-$0.43 & 0.12 & $-$1.43 & $-$0.59 & 0.07 & $-$1.95    &  & 15.19  & 23.24  & 15.42  \\
119 & IV & 1345.05  & 0.19  & $-$179.65  & 0.10  & $-$4.53    & 0.03 & 0.11 & 0.10 & $-$0.41 & 0.06 & $-$1.35   &  & 2.41  & 6.73  & 4.53  \\
120 & IV & 1345.68  & 0.07  & $-$180.35  & 0.05  & $-$4.18    & $-$1.16 & 0.06 & $-$3.86 & 0.04 & 0.04 & 0.12   &  & 30.94  & 29.77  & 19.45  \\
121 & IV & 1345.75  & 0.13  & $-$175.55  & 0.07  & $-$4.18    & $\cdots$ & $\cdots$ & $\cdots$ & $\cdots$ & $\cdots$ & $\cdots$ &  & $\cdots$ & $\cdots$ & 17.06  \\
122 & IV & 1346.17  & 0.18  & $-$181.41  & 0.10  & $-$3.65    & $\cdots$ & $\cdots$ & $\cdots$ & $\cdots$ & $\cdots$ & $\cdots$ &  & 2.13  & $\cdots$ & $\cdots$ \\
123 & IV & 1346.58  & 0.06  & $-$180.29  & 0.04  & $-$4.01    & $-$0.98 & 0.09 & $-$3.24 & $-$0.66 & 0.06 & $-$2.20   &  & 28.78  & 13.40  & 13.95  \\
124 & IV & 1347.64  & 0.09  & $-$180.68  & 0.05  & $-$3.83    & $\cdots$ & $\cdots$ & $\cdots$ & $\cdots$ & $\cdots$ & $\cdots$ &  & 11.53  & $\cdots$ & $\cdots$ \\
125 & IV & 1352.70  & 0.12  & $-$178.68  & 0.07  & $-$4.01    & $\cdots$ & $\cdots$ & $\cdots$ & $\cdots$ & $\cdots$ & $\cdots$  &  & 16.28  & 6.18  & $\cdots$ \\
126 & IV & 1352.85  & 0.20  & $-$178.36  & 0.10  & $-$4.18    & $-$0.76 & 0.16 & $-$2.51 & $-$1.13 & 0.08 & $-$3.74 &  & 12.43  & 10.02  & 9.21  \\
127 & IV & 1353.82  & 0.15  & $-$179.17  & 0.09  & $-$3.83    & $\cdots$ & $\cdots$ & $\cdots$ & $\cdots$ & $\cdots$ & $\cdots$ &  & 6.22  & $\cdots$ & $\cdots$ \\
128 & IV & 1354.54  & 0.48  & $-$184.84  & 0.43  & $-$4.53    & $\cdots$ & $\cdots$ & $\cdots$ & $\cdots$ & $\cdots$ & $\cdots$ &  & $\cdots$ & 1.43  & $\cdots$ \\
129 & IV & 1354.72  & 0.49  & $-$178.35  & 0.26  & $-$4.36   & $\cdots$ & $\cdots$ & $\cdots$ & $\cdots$ & $\cdots$ & $\cdots$  &  & 3.52  & 5.46  & $\cdots$ \\
130 & IV & 1361.74  & 0.15  & $-$164.10  & 0.09  & $-$3.65   & $\cdots$ & $\cdots$ & $\cdots$ & $\cdots$ & $\cdots$ & $\cdots$ &  & 2.52  & $\cdots$ & $\cdots$ \\
131 & IV & 1362.97  & 0.18  & $-$163.11  & 0.10  & $-$3.83   & $\cdots$ & $\cdots$ & $\cdots$ & $\cdots$ & $\cdots$ & $\cdots$ &  & 4.36  & $\cdots$ & $\cdots$ \\
\end{longtable}
\end{landscape}
}
% end of onltab

%end of the main text

\end{document}